\documentclass[]{aa} 

\usepackage{graphicx}
\usepackage{txfonts}
\usepackage{natbib}
\usepackage{longtable}

\begin{document}

\title{ 
The degeneracy between dust colour temperature and spectral index
}
\subtitle{Comparison of methods for estimating the $\beta(T)$ relation}

\author{M.     Juvela\inst{1},
        J.     Montillaud\inst{1},
        N.     Ysard\inst{2},
        T.     Lunttila\inst{1}
}

\institute{
Department of Physics, P.O.Box 64, FI-00014, University of Helsinki,
Finland, {\em mika.juvela@helsinki.fi}
\and
Institut d’Astrophysique Spatiale, UMR 8617, CNRS/Universit\'e
Paris-Sud 11, 91405 Orsay, France
}

\authorrunning{M. Juvela et al.}

\date{Received September 15, 1996; accepted March 16, 1997}

\abstract
{
Sub-millimetre dust emission provides information on the physics of
interstellar clouds and dust. However, noise can produce arteficial
anticorrelation between the colour temperature $T_{\rm C}$ and the
spectral index $\beta$. These artefacts must be separated from the
intrinsic $\beta(T)$ relation of dust emission.
}
{
We compare methods that can be used to analyse the $\beta(T)$
relation. We wish to quantify their accuracy and bias, especially for
observations similar to those made with the $Planck$ and $Herschel$.
}
{
We examine sub-millimetre observations that are simulated either as simple
modified black body emission or using 3D radiative transfer modelling. We use
different methods to recover the ($T$, $\beta$) values of individual objects and
the parameters of the $\beta(T)$ relation. In addition to $\chi^2$ fitting, we
examine the results of the SIMEX method, basic Bayesian model, hierarchical
models, and a method that explicitly assumes a functional form for $\beta(T)$.
The methods also are applied to one field observed by {\it Herschel}.
}
{
All methods exhibit some bias, even in the idealised case of white noise. The
results of the Bayesian method show significantly lower bias than direct $\chi^2$
fits. The same is true for hierarchical models that also result in a smaller
scatter in the temperature and spectral index values. However, significant bias
was observed in cases with high noise levels. When the signal-to-noise ratios are
different for different sources, all $\beta$ and $T$ estimates of the
hierarchical model are biased towards the relation determined by the data with
the highest S/N ratio. This can significantly alter the recovered $\beta(T)$
function.  With the method where we explicitly assume a functional form for the
$\beta(T)$ relation, the bias is similar to the Bayesian method. In the case of
the actual $Herschel$ field, all methods agree on some degree of
anticorrelation between $T$ and $\beta$.
}
{
The Bayesian method and the hierarchical models can both reduce the noise-induced
parameter correlations. However, all methods can exhibit non-negligible bias.
This is particularly true for hierarchical models and observations of varying
signal-to-noise ratios and this must be taken into account when interpreting the
results.
}

\maketitle
%

\section{Introduction}

Observations of sub-millimetre and millimetre dust emission are
considered one of the most reliable ways of mapping dense clouds and
cloud cores in different phases of the star formation process
\citep{Motte1998, Andre2000, Enoch2007, Andre2010}. Many other tracers
suffer from some serious problems, especially in the case of dense
pre-stellar and protostellar cores. The interpretation of molecular
line data is complicated by the abundance variations and the optical
depth effects and very high-resolution maps of dust extinction and
scattering are difficult to obtain \citep{Lombardi2006, Goodman2009,
Juvela2008, Juvela2009}.

In addition to being a tracer of the structure of interstellar clouds,
the dust emission carries information on the properties of the dust
grains themselves. These are expected to vary from region to region.
The grain optical properties depend on their chemical composition 
and the dust size
distribution is modified by the competing effects of dust erosion
(e.g., by shock waves) and aggregation processes
\citep[e.g.,][]{Ormel2011, Jones2011}. The grain emissivity may depend
even on the temperature of the grains but is certainly modified by the ice
mantles that are acquired in cold cloud cores
\citep{Ossenkopf1994,Stepnik2003, Meny2007, Compiegne2011}. Because
these changes depend on the local current and past
conditions, they could, in principle, be used to trace the physical
evolution of the clouds and their dust during the star-formation
process. Thus, accurate measurements of the dust properties would be
very valuable.

Unfortunately, it is difficult to deduce intrinsic dust properties
based on the observed radiation. Some of the difficulties are caused
by the line-of-sight temperature variations that broaden the emission
spectrum, decreasing the observed spectral index $\beta$ 
\citep{Shetty2009a, Malinen2011, Juvela2012_TB}. At the same time,
the dust colour temperature $T_{\rm C}$ overestimates the mass
averaged physical dust temperature, and the total dust mass becomes
underestimated \citep{Evans2001, StamatellosWhitworth2003,
Malinen2011, YsardJuvela2012}. If the exact nature of the
line-of-sight temperature variations is not known, it is 
difficult to accurately measure the intrinsic dust emissivity spectral
index. 

We want to know, to what accuracy the basic parameters of dust temperature and
dust emissivity spectral index can be deduced from observations. In the
present paper, we concentrate on the complications caused by the noise present in
the measured sub-millimetre intensity values. When dust observations are fitted
with modified black body spectra, the dust colour temperature, $T_{\rm C}$, and the
observed dust spectral index, $\beta$, are partly degenerate. The noise will
scatter the observed $T_{\rm C}$ and $\beta$ values over an elongated ellipse that
shows apparent negative correlation between the parameters. In the limit of low
noise, the errors are correlated but the expectation values are not significantly
biased. If the noise is larger, error regions become banana-shaped because the
$\beta(T_{\rm C})$ relation gets steeper at lower temperatures \citep{Shetty2009a,
Shetty2009b, Veneziani2010, Paradis2010, GCC_II}. The parameter distributions
can become skew and, in rare cases, the $\chi^2$ values of the modified black body
fits may even have separate minima that are several degrees apart in $T_{\rm C}$
\citep{Juvela2012_chi2}. Under these conditions, the ($T_{\rm C}$, $\beta$) values
determined for individual sources (or for pixels in a map) are not only uncertain
but also biased.

The $\beta(T_{\rm C})$ relation induced by the noise is usually
steeper than the average $\beta(T_{\rm C})$ relation derived from
observations of a set of sources \citep{Dupac2003, Desert2008,
PlanckI}. Therefore, the noise tends to make the estimated
$\beta(T_{\rm C})$ relation steeper than the true relation. The bias
can be significant even at noise levels below 10\%. With
perfect knowledge of the observational noise, one should be able to
eliminate these effects and to recover the unbiased ($T_{\rm C}$, $\beta$)
estimates for the sources and to determine the true overall
$\beta(T_{\rm C})$ relation. Note that even this would be the true
relation only in the sense that it would accurately describe the
observed radiation. It would still not describe the intrinsic grain
properties because of line-of-sight temperature variations within the
source and, in extreme cases, by opacity effects. 

The effect of the noise can be tackled in different ways. 
In the Bayesian formulation one is probing the posterior probability
$P(\theta|d) \propto P(\theta) P(d|\theta)$, where $P(\theta)$ is the {\it a priori} probability
of the parameters $\theta$ and $P(d|\theta)$ the likelihood function. The
probability $P(\theta|d)$ can be estimated with Markov Chain Monte Carlo 
calculations \citep[e.g.][]{Veneziani2010} to recover less biased
$\beta$ and $T$ estimates.
In the hierarchical Bayesian method, one additionally includes a
global model for the distribution of the $T$ and $\beta$ (and column
density) values \citep{Kelly2012}. 
If one is interested mainly in the overall relation $\beta(T_{\rm
C})$, one also can start with the observed values and employ
direct Monte Carlo simulation to estimate how much the estimated
$\beta(T)$ relation is affected by the assumed level of noise
\citep{PlanckI}. The procedure also can be automatised, looking for a
$\beta(T)$ relation that corresponds to the observations when the
effect of the noise, estimated with Monte Carlo methods, is added.
On the other hand, SIMEX \citep{CookStefanski1994} is a very general
method where the results are examined as a function of the input noise
and the final de-biased estimate is obtained by an extrapolation to
zero noise level.
In this paper we compare the methods listed above, also including some
variations of the hierarchical models. Our main goal is to quantify
the accuracy by which the $\beta(T_{\rm C})$ relation can be derived,
for example, from observations carried out by {\it Planck}
\citep{Tauber2010} and {\it Herschel} \citep{Pilbratt2010}. This
includes especially the examination of the bias.

The content of the paper is the following. In Sect.~\ref{sect:methods} we present the methods that
are used to analyse intensity measurements and to derive the parameters of $\beta(T_{\rm C})$
relation. Section~\ref{sect:data} describes the data used in the tests. These include simple
modified black body spectra (Sect.~\ref{sect:simu_data}), surface brightness maps obtained with radiative
transfer modelling (Sect.~\ref{sect:MHD_model}), and actual {\it
Herschel} data
(Sect.~\ref{sect:Herschel_data}). The corresponding results are presented in
Sections~\ref{sect:results_mbb}--\ref{sect:results_Herschel}. The results are discussed in
Sect.~\ref{sect:discussion} and the final conclusions are listed in Sect.~\ref{sect:conclusions}.


\section{Analysis methods} \label{sect:methods}

In this section we present the methods that are used to estimate the
($T$, $\beta$) values of individual sources and the parameters ($A$,
$B$) of the $\beta(T)$ relation (see Eq.\ref{eq:AB}) for a collection
of sources.

\subsection{The $\chi^2$ fitting}

The derivation of the modified black body parameters and of the
$\beta(T)$ relation is often based on direct maximum likelihood
estimates. Apart from possible filtering of outliers and other
allowances for deviations from the normal statistics, the least
squares fitting still remains the basic method. The analysis is
performed in two stages. The ($T$, $\beta$) values are first estimated
for each source (or pixel) separately, fitting the intensities at the
observed wavelengths with a modified black body law, using weighted
least squares. The estimated ($T$, $\beta$) points are then fitted
with a parametric curve that, in our case, is Eq.~\ref{eq:AB} with the
parameters $A$ and $B$.

We call the above $\chi^2$ method because both steps steps can be
based on $\chi^2$ minimisation. The uncertainties $\delta T$ and
$\delta \beta$ can be estimated based on the measurement errors using,
for example, Monte Carlo methods. When the analytical $\beta(T)$
relation is fitted, the uncertainties of the individual intensity
measurements are no longer considered and one only relies on the
previously derived $\delta T$ and $\delta \beta$ estimates. We
explicitly ignore the information on the correlations of the $T$ and
$\beta$ errors.

\subsection{SIMEX method}

In the SIMEX method \citep{CookStefanski1994}, one examines how the
results change as a function of the noise. The final result
is obtained by extrapolating this relation to a zero noise level.
Because the ($T$, $\beta$) measurements of individual sources usually do not
exhibit clear bias, SIMEX is not useful in their analysis.
Only when the signal-to-noise ratio is very low, the error
distributions of, e.g., the spectral index can become skew
\citep{Juvela2012_chi2} and SIMEX could be used to quantify the
effect. In this paper, we only examine SIMEX estimates of
the $A$ and $B$ parameters of the $\beta(T)$ relation.

In each case, we generate 1000 noise realisations where the noise
level is varied from that of the actual observation to a case with
three times the original noise. We add Gaussian noise to the intensity
measurements (that already contain the original observational noise),
perform the modified black body fits for all sources, and fit the
$\beta(T)$ relation with the least squares method. This gives us 1000
points describing the change of the parameters $A$ and $B$ as
functions of the noise. We perform linear fits to these relations and
the extrapolation to zero noise level gives the bias-corrected
estimates of $A$ and $B$.

The use of the SIMEX method requires reliable estimates of the noise
level of the data. Because of the extrapolation, the statistical noise
will be enhanced but, if the noise dependence is well defined, SIMEX
could reduce the bias. There is no guarantee that the functional form
of the noise dependence is similar at low and high noise levels. The
result will depend on the range of noise levels examined and on the
functional form used for the fitting and extrapolation. If the
extrapolation function is not adequate for the problem at hand, the
method could even increase the bias.

\subsection{Bayesian method} \label{sect:BM}

In the Bayesian method one
estimates the posterior probability $p(\theta|d)$ of model parameters
$\theta$ for given data $d$. It presents the advantage of giving the
parameters that maximise $p(\theta|d)$ and, at the same time,
providing error estimates via their probability distribution.
Using Bayes formula, the posterior probability of model parameters for
given data, $p(\theta|d)$ can be written as
\begin{equation}
p(\theta|d) \propto p(\theta) \, p(d|\theta).
\label{eq:bayes}
\end{equation}
Here $p(\theta)$ is the prior probability assigned to the parameters $\theta$ and
the last term, $p(d|\theta)$, is called the likelihood function that gives the
probability of data $d$ for a given set of model parameters $\theta$. The parameter
vector $\theta$ contains estimates for the intensity (or flux), temperature, and
spectral index of each source.  In our case $p(d|\theta)$ corresponds to combined
probability of the fits of modified black body spectra to the observations of
individual sources. We assume a flat prior. Thus, the probability can be
calculated based on the combined $\chi^2$ value of those fits, assuming normal
statistics
\begin{equation}
\ln p(d|\theta) \propto 
\sum_{i=1}^{N_{\rm S}}
\sum_{j=1}^{N_{\rm f}}
\left[  
\frac{S_{i,j}-S_{\rm model}}{dS_{i, j}},
\right]^2
\end{equation}
where the sum includes flux measurements $S$ and their error estimates $dS$ for all the $N_{\rm
S}$ sources and $N_{\rm f}$ frequencies.
We evaluate $p(\theta|d)$ using Markov Chains Monte Carlo (MCMC).
In the Metropolis-Hastings algorithm, one
generates parameter steps $\theta_{\rm i} \rightarrow \theta_{\rm
{i+1}}$, using some propositional probability distribution for the
step. The step is then accepted if 
\begin{equation}
\frac{p(\theta_{\rm i+1}|d)}{p(\theta_{\rm i}|d)} > u,
\end{equation}
where $u$ is a uniformly distributed random number between 0 and 1.
Thus, the calculation only requires the knowledge of the ratio of the
probabilities. The accepted $\theta_{\rm i}$ values sample the
posterior probability distribution $p(\theta|d)$ \citep[see,
e.g.,][]{Veneziani2010}.

We generate the parameter steps using normal distribution. The
relative step sizes of the different parameters are fixed beforehand
but the absolute step length is adjusted with a single multiplicative
factor to keep the acceptance rate close to 20\%. After the initial
burn-in phase, MCMC samples the posterior probability distribution
Eq.~\ref{eq:bayes} and, using values from a sufficiently long piece of
the chain, one can directly estimate the distributions of the
parameters. We run the MCMC in pieces of about million steps. The
parameter distributions in each piece of the chain are examined to
make sure when a convergence was reached. The reported results are
calculated from the last part of the chain.

It would be possible to extract samples from the MCMC parameter
distributions to estimate distributions also for the parameters $A$
and $B$ of the $\beta(T)$ relation. However, we fit the $\beta(T)$
relation to the average $T$ and $\beta$ values obtained for each
individual source. This allows direct comparison with other methods
(e.g., $\chi^2$ and SIMEX).

\subsection{Bayesian hierarchical model} \label{sect:HM}

In the Bayesian hierarchical model one combines the models of the individual
sources with a global model describing the overall distribution of source
parameters \citep{Kelly2012}. In our case, the basic source parameters are the
intensity $I_{i}$ at a reference wavelength, the temperature $T_{i}$, and the
spectral index $\beta_{i}$. When analysing maps, each pixel is considered as a
separate source.

The posterior probability is calculated from
\begin{equation}
p(I, T, \beta, \delta | S) \propto 
p(\theta)  \prod_{i=1}^{N_{\rm S}}
\left[
p(I_i, T_i, \beta_i|\theta)
\prod_{j=1}^{N_{\rm f}} p(S_{i,j}| I_i, T_i, \beta_i, \delta_j)
\right].
\label{eq:hm}
\end{equation}
The term $p(\theta)$ denotes the {\it a priori} probability taking into account all parameters
included in the model.
The formula includes the product over $N_{\rm S}$ sources. Inside the
square brackets, the first term is the probability of the parameters
($I_i$, $T_i$, $\beta_{i}$) of the source $i$, given the underlying
parameter vector $\theta$. 
In addition to the intensity, temperature, and spectral index
values of individual sources, $\theta$ contains parameters describing
the global probability distribution of the $I_i$, $T_i$, and $\beta_i$
values (in practice, their expectation values and a covariance matrix)
and optionally parameters describing calibration errors (see below).
The second term in the square brackets is the probability of the data
that is calculated for given modified black body parameters and for
$N_{\rm f}$ observed frequencies. The values $S_{i,j}$ are the
observed intensities (or fluxes) for source with index $i$ and for
frequency with index $j$. The parameters $\delta_j$ are used to
account for possible calibration error (see below).
Equation~\ref{eq:hm} corresponds to Eq.~(9) in \citet{Kelly2012}. 

In Eq.~\ref{eq:hm} one must calculate, firstly, the probabilities of
the individual measurements and, secondly, the probabilities of
the parameters of the fitted modified black bodies. Both calculations
can be done based on different assumptions of the statistics. If we
assume the errors to be normally distributed, the first probability is
simply
\begin{equation}
p(S_{i,j}|I_i, T_i, \beta_i, \delta_j) \propto
{\rm exp}\left[ 
-\frac{1}{2} \left(\frac{S_{i,j}- \delta_j S_j(I_i, T_i, \beta_i)}{\sigma_{i,j}}\right)^2
\right],
\label{eq:normal1}
\end{equation}
where $S_j(I_i, T_i, \beta_i)$ is the intensity for a modified black
body with parameters $I_i, T_i, \beta_i$ and frequency $j$. This leads
to normal weighted least squares.  In the above equation we also
include parameters $\delta_{i,j}$ that can be used in MCMC
calculations to account for calibration errors. These are ``nuisance''
parameters that are part of $\theta$ but  are marginalised out
from the final results. Following \citet{Kelly2012}, the probability
can alternatively be calculated using Student distribution as
\begin{equation}
p(S_{i,j}|I_i, T_i, \beta_i, \delta_j) \propto
\frac{1}{\sigma_{i,j}}
\left[
1 + \frac{1}{d}
\left( 
\frac{S_{i,j} - \delta_{j} S_{j}(I_i, T_i, \beta_i)}{\sigma_{i,j}}\right)^2
\right]^{-(d+1)/2}.
\label{eq:student1}
\end{equation}
We use the formula with $d=3$ degrees of freedom. 

The term $p(I_i, T_i, \beta_i|\theta)$ also can be calculated using Student
distribution
\begin{equation}
p(I_i, T_i, \beta_i|\theta) \propto
\frac{1}{ |\Sigma|^{1/2}}
\left[
1 + \frac{1}{d}  (x_i-\mu)^T \Sigma^{-1} (x_i-\mu)
\right]^{-(d+3)/2},
\label{eq:student2}
\end{equation}
where $x_i=(I_i, T_i, \beta_i)$, and the parameters $\theta$ consist of
$\mu$ containing the expectation values of ($I$, $T$, $\beta$) and the
covariance matrix $\Sigma$. Following \cite{Kelly2012}, the degrees of freedom
is set to the value of $d=8$.
Alternatively, we can again assume normal distribution. Because the
matrix $\Sigma$ is not constant, the calculation does not reduce to
the estimation of the $\chi^2$ value. Instead, we must use the full
formula
\begin{equation}
p(I_i, T_i, \beta_i|\theta) \propto
\frac{1}{|\Sigma|^{1/2}}
\exp \left[ -0.5 \times \left[ (x_i-\mu)^T \Sigma^{-1} (x_i-\mu)
\right] \right].
\label{eq:normal2}
\end{equation}
In both cases, $\theta$ includes parameters for the expectation
value and covariances of $I_i$, $T_i$, and $\beta_i$. 

The solution is searched with the Markov Chain Monte Carlo (MCMC)
method, using the Metropolis-Hastings algorithm. We have a long
parameter vector that contains the parameters of the individual
sources (i.e., $I_{i}$, $T_{i}$, $\beta_{i}$) and global
parameters consisting of the elements of the $\mu$ vector and the
$\Sigma$ matrix.  In practice, $\Sigma$ is defined by the three
standard deviations and the correlations of $I_{i}$, $T_{i}$,
$\beta_{i}$. These parameters form part of the MCMC parameter vector,
with the additional constraint that the resulting covariance matrix
must be positive-definite.
In the following, we refer to hierarchical models as HM or
specifically as HM(N) or HM(S), depending on whether the calculations
are based on Eqs.~\ref{eq:normal1} and \ref{eq:normal2} or
Eqs.~\ref{eq:student1} and \ref{eq:student2}, respectively.

\subsection{Model with an explicit $\beta(T_{\rm C})$ law}
\label{sect:MC}

One can use Monte Carlo simulations to determine what effect an assumed
level of noise has on the estimated $\beta(T)$ law. This also can be done in an
iterative fashion. One constructs a model consisting of the values ($I_i$, $T_i$,
$\beta_i$) of the sources, carries out the Monte Carlo simulation to estimate the
observational bias, and one optimises the match between the model predictions
(including effects of noise) and the actual observations.

We use this idea, combining it with the assumption that all sources follow a
single $\beta(T)$ relation that can be described by Eq.~\ref{eq:AB}. In the
following, this is referred to as the MC method. Because $T_i$ together with $A$
and $B$ determines a unique value of $\beta_i$, the full MCMC parameter vector
only contains the values of $A$ and $B$ and the $I_{i}$ and $T_{i}$ of the
individual sources. The match between this model and the observations is handled
through a separate Monte Carlo simulation step. The current MCMC parameters
define a modified black body spectrum for each source. Based on these model
spectra and the assumed noise level, we carry out a simulation where we add noise
to the measurements and re-calculate the modified black body fits for each source
keeping both $T$ and $\beta$ as free parameters. This gives us samples of the
apparent $(T^{\rm MC}, \beta^{\rm MC})$ values that are affected by the
noise. Using these realisations of ($T_i^{\rm MC}$, $\beta_i^{\rm MC}$) values
for all sources, we can perform a fit to estimate the $A^{\rm MC}$ and $B^{\rm
MC}$ values that also are biased by noise. If the original MCMC parameters are
correct, the bias in the $A^{\rm MC}$ and $B^{\rm MC}$ values should be identical
to the bias in the actual observations. Thus we impose a constraint that the
$A^{\rm MC}$ and $B^{\rm MC}$ values must be the same as the values $A^{\rm Obs}$
and $B^{\rm Obs}$ that are derived directly from the observations. The
constraint is added to the MCMC procedure by including in the probabilities an
additional term describing the match between ($A^{\rm MC}$, $B^{\rm MC}$) and
($A^{\rm Obs}$, $B^{\rm Obs}$). In our implementation, these probabilities are
calculated from normal distribution using $\sigma(A)=0.05$ and $\sigma(B)=0.035$.

One major difference to the previous hierarchical model is the {\it a priori}
assumption of a single $\beta(T)$ relation of the given form. If this assumption
is not correct, this could be recognised only from higher $\chi^2$ values in the
fits of the individual sources. In practice, this may be difficult because the
noise is usually not known to very high accuracy. On the other hand, if the
$\beta(T)$ relation can be described by the assumed formula, the results should
have low statistical noise because of the strong additional constraint. Another
significant difference is the explicit use of Monte Carlo simulation to describe
the effects of the noise. This is more general than Eq.~\ref{eq:hm}, where one
also has to choose a model for the global distribution of ($I_{i}$, $T_{i}$,
$\beta_{i}$) values. In particular, the Monte Carlo simulation automatically
takes into account the possible asymmetry and curvature of the ($T$, $\beta$)
error regions and how these affect the apparent $\beta(T)$ relation.

\begin{table}
\caption{The methods used in the estimation of the $\beta(T)$ relation.}
\centering
\begin{tabular}{lll}
\hline\hline
Name                 &  Description \\
\hline
$\chi^2$             &  Least squares fits of modified black bodies \\
                     &  and of the $\beta(T)$ relation \\
SIMEX                &  The SIMEX method, extrapolation of the \\
                     &  $A$ and $B$ parameters to zero noise level \\
MC                   &  Model with explicit functional form of $\beta(T)$, \\
                     &  with noise bias estimated with Monte Carlo \\
BM                   &  Bayesian model, Markov Chain Monte Carlo estimates \\                     
HM(S)                &  Hierarchical model with Student distributions,  \\
                     &  Eqs.~\ref{eq:student1} and \ref{eq:student2}, 
                        as in \citet{Kelly2012} \\
                     &  (without parameters $\delta$) \\
HM(N)                &  As HM(S) but using normal distributions,  \\
                     &  Eqs.~\ref{eq:normal1} and \ref{eq:normal2}  \\
HMC                  &  Hierarchical model including $\delta$ calibration \\
                     &  parameters (otherwise usually as HM(N)) \\
\hline
\end{tabular}
\label{table:methods}
\end{table}

\begin{table}
\caption{The limits of the priors used in MCMC calculations and in
hierarchical models. Apart from $\delta_j$ (see Sect.~\ref{sect:MHD_noise}), the
priors are flat. The parameters not listed in the table have flat priors without
limits.} 
\centering
\begin{tabular}{llll}
\hline \hline
Parameter           &   Min    & Max         &  Note \\
\hline
$\mu_1$=$<T>$       &  5.0     &     30.0    & expectation value of temperature\\
$\mu_2$=$<\beta>$   &  0.0     &      4.5    & expectation value of spectral index\\
$I_i$               &  0.1     &     10.0    & intensity of a source/pixel \\
                    &          &             & at a reference wavelength \\
                    &          &             & (no upper limit in the MHD model)\\
$T_i$               &  5.0     &     30.0    & temperature of a source/pixel\\
$\beta_i$           & -6.0     &     +6.0    & spectral index in modified black body tests \\ 
$\beta_i$           &  0.0     &     +4.5    & spectral index in calculations with MHD \\
                    &          &             & models and Herschel observations \\ 
$\delta_j$          &  0.3     &      2.5    & factors for calibration errors\\  
\hline
\end{tabular}
\label{table:priors}
\end{table}


\section{Analysed observations} \label{sect:data}

\subsection{Simulated modified black body spectra}  \label{sect:simu_data}

In this paper we are only interested in the effects of observational
noise and, therefore, choose for the noiseless data a model that is as
simple as possible. The data consist of intensity (or flux) values
that, for each source, are calculated according to a single modified
black body law
\begin{equation}
I_{\nu} \propto B_{\nu}(T) \times \nu^{\beta},
\end{equation}
where $\beta$ is a fixed emissivity spectral index and $B_{\nu}(T)$ is
the Planck radiation law for temperature $T$. The parameters ($T$,
$\beta$) merely describe the properties of the observed radiation.  In
the case of real observations, these would be called the colour
temperature and the apparent spectral index to emphasise the
difference to the actual physical temperature and emission properties
of the dust grains. In the rest of the paper, we refer to the colour
temperature simply with the symbol $T$. We add to the calculated
intensities pure Gaussian, uncorrelated noise. In the first tests the
relative noise is fixed to either 5\% or 10\% of the original
noiseless signal. We use two combinations of wavelengths. The first
one consists of 100.0, 160.0, 250.0, 350.0, and 500.0\,$\mu$m, in
correspondence to a possible set of wavelengths observed with {\it
Herschel}. The second set is 100.0, 350.0, 550.0, and 850.0\,$\mu$m.
These wavelengths were used in \citet{PlanckI} where the data
consisted of the IRAS 100\,$\mu$m maps and the three highest frequency
bands of the {\it Planck} satellite.

We analyse observations of a small number of sources. Apart from the
effects of noise, the ($T$, $\beta$) values of the sources are assumed
to follow a single $\beta(T)$ relation. For this relation we have
selected a functional form
\begin{equation}
\beta(T) =  A\,\, (T/20.0\,{\rm K})^{B}
\label{eq:AB}
\end{equation}
\citep[see, e.g.,][]{Desert2008, Paradis2010, PlanckI}.  For the
parameter $B$, we use the values of -0.3, 0.0, or +0.3. A negative
value of $B$ seems more consistent with most observations but we also
examine a flat relation ($B$=0.0) and a positive $T$-$\beta$
correlation ($B$=+0.3) to see the behaviour of the analysis methods in
these different situations. To be roughly consistent with data on very
dense clouds (for any of the chosen values of $B$), the parameter $A$
is fixed to a value of 2.2. 
In most tests, we use a sample of 50 sources with temperatures
distributed regularly between 8.0\,K and 20.0\,K. To build statistics,
we typically examine 40 noise realisations of each set of
observations.

\subsection{Radiative transfer model} \label{sect:MHD_model}

As a more realistic set of simulated observation, we use surface brightness maps that
are calculated with three-dimensional radiative transfer modelling, using a density field obtained from
a magnetohydrodynamical (MHD) simulation.

The surface brightness maps are based on the MHD simulations of \citet{PadoanNordlund2011} that were
carried out on a grid of 1000$^3$ cells. For radiative transfer modelling the density field was
regridded onto a hierarchical grid \citep[see][]{Lunttila2012}, the temperature distribution of the
large grains was solved. The calculations resulted in 1000$\times$1000 pixel surface brightness
images. The dust model corresponded to normal Milky Way diffuse interstellar dust ($R_{V}$=3.1) as
described in \citet{Draine2003}. The MHD model and the radiative transfer calculations have been
described in detail in \citet{Juvela2012_mhdfil}. There original cloud corresponded to a linear size
of 6\,pc and a mean hydrogen density of $\sim$200\,cm$^{-3}$ but only the cloud opacity is important
for the modelling of dust emission. For the present study we selected a smaller region that covers
approximately one quarter of the projected area of the full model. The surface brightness images at
160\,$\mu$m, 250\,$\mu$m, 350\,$\mu$m, and 500\,$\mu$m were down-sampled to 101$\times$101 pixel
images with an assumed pixel size of 6$\arcsec$. We added to the images noise using as a starting
point noise values of 3.7, 1.20, 0.85, and 0.35\,MJy\,sr$^{-1}$ per {\it Herschel} beam, assuming
beam sizes 12.0$\arcsec$, 18.3$\arcsec$, 24.9$\arcsec$, and 36.3$\arcsec$, respectively. The noise
level was adjusted with parameter $N$ in the range of 0.0-1.0 times the default values.  For
analysis the maps were convolved to a common resolution of 40$\arcsec$.

\subsection{Herschel observations} \label{sect:Herschel_data}

The final tests will be carried out using actual {\it Herschel} data at 160\,$\mu$m, 250\,$\mu$m,
350\,$\mu$m, and 500\,$\mu$m. The observations are for the field G109.80+2.70 that was analysed in
\cite{GCC_II} using the $\chi^2$ method (the target PCC288 in that paper). The data used in this
paper are not identical. The observations of the SPIRE instrument, 250\,$\mu$m-500\,$\mu$m, have
been reprocessed using the Herschel Interactive Processing Environment HIPE v.9.0\footnote{HIPE is a
joint development by the Herschel Science Ground Segment Consortium, consisting of ESA, the NASA
Herschel Science Center, and the HIFI, PACS and SPIRE consortia}, with specific processing to
improve the subtraction of the scan baselines. The SPIRE maps are the product of direct projection
onto the sky and averaging of the time ordered data. The 160\,$\mu$m data of the PACS instrument
have been processed with HIPE and the final map has been produced with the Scanamorphos program
\citep{Roussel2012}. The PACS map still suffers from some artefacts. The surface brightness appears
to rise too much towards the northern corner of the map and, according to comparison with IRAS and
AKARI data of the region, the 160\,$\mu$m map also suffers from a small South-North gradient.
Instead of correcting or masking the artefacts, we use the data directly in our comparison of the
analysis methods. The zero point of the surface brightness scale has been estimated with the help of
{\it Planck} and IRAS data \citep[see][]{GCC_II, GCC_III} and all data were convolved
to $40\arcsec$ before the analysis. The noise levels were estimated by selecting a flat region in
the northern part of the map and by calculating the standard deviation between the observations and
the same map convolved to a resolution of 2$\arcmin$. The noise per a 40$\arcsec$ beam was found to
be 6.75, 4.34, 2.30, and 1.21\,MJy\,sr$^{-1}$, for the four bands in order of increasing wavelength.
After masking areas outside and near the boundaries of the PACS map, we are left with an area of
$\sim 10\arcmin \times 10\arcmin$. When sampled at 12$\arcsec$ intervals, the data set consists of
$\sim$7850 pixels at each wavelength.


\section{Results for idealised modified black body spectra} \label{sect:results_mbb}

In this section we examine a series of simulated observations where, apart
from the noise, each observed spectrum exactly follows the assumed model of a
modified black body. We compare the results for the methods listed in
Table~\ref{table:methods} but do not yet include calibration factors $\delta$ in
the calculations with the HM method. In Sect.~\ref{sect:rnoise}, we use
observations with fixed relative uncertainty for all bands and sources. In
Sect.~\ref{sect:Tsplit} we study the effect of analysing the low and high
temperature sources separately. We then move to cases where the absolute
uncertainty is fixed, emulating more closely data from the current {\it Herschel}
and {\it Planck} observations (Sect.~\ref{sect:anoise}).
In Sect.~\ref{sect:number} we check the effect of the size of the source sample
before, in Sect.~\ref{sect:NT}, examining a case where the source temperatures are
distributed according to normal distribution rather than uniformly between 8\,K and
20\,K.

\subsection{Observations with fixed relative error} \label{sect:rnoise}

\begin{figure}
\centering
\includegraphics[width=8.7cm]{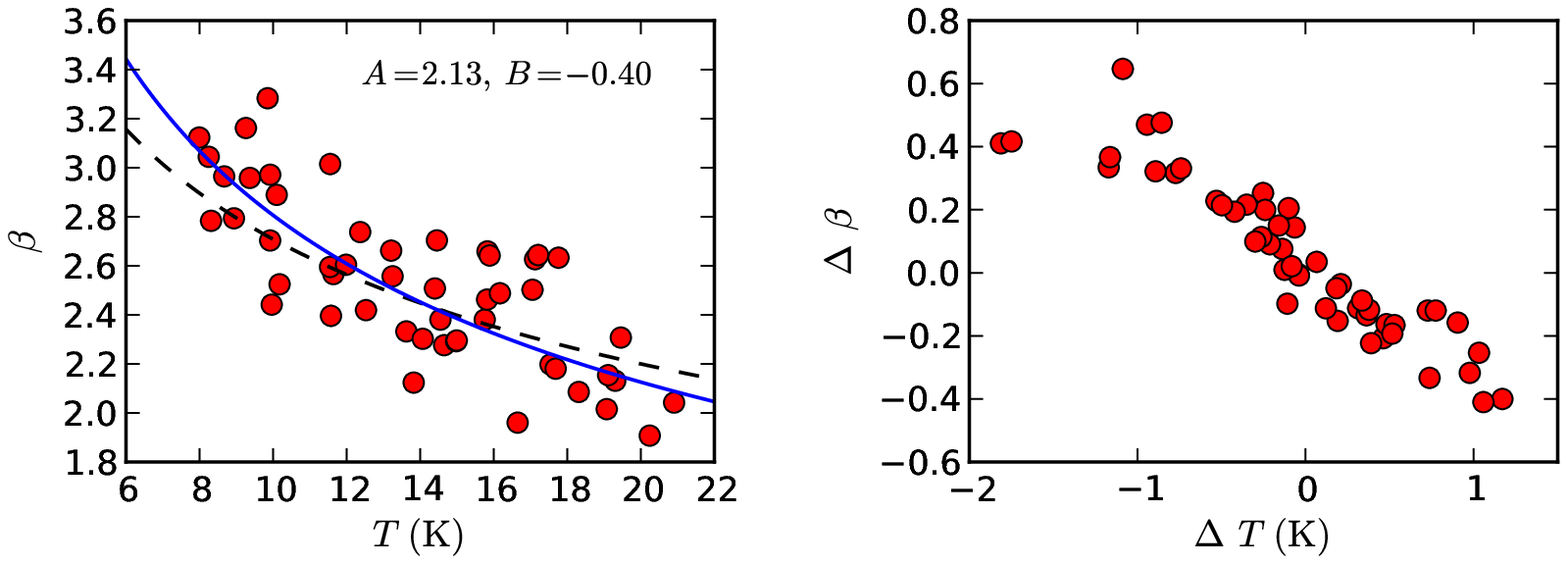}
\caption{
The result of the $\chi^2$ method for the case where the sources
follow a $\beta(T)$ relation with parameters $A=2.2$ and $B=-0.3$ and
the measurements in {\it Herschel} bands have a relative noise of
5\%. The left hand frame shows the estimated $T$ and $\beta$ values
and the fitted $\beta(T)$ relation. The right hand frame shows the
absolute errors in the $T$ and $\beta$ values of the sources.
}
\label{fig:chi2}%
\end{figure}

\begin{figure}
\centering
\includegraphics[width=8.7cm]{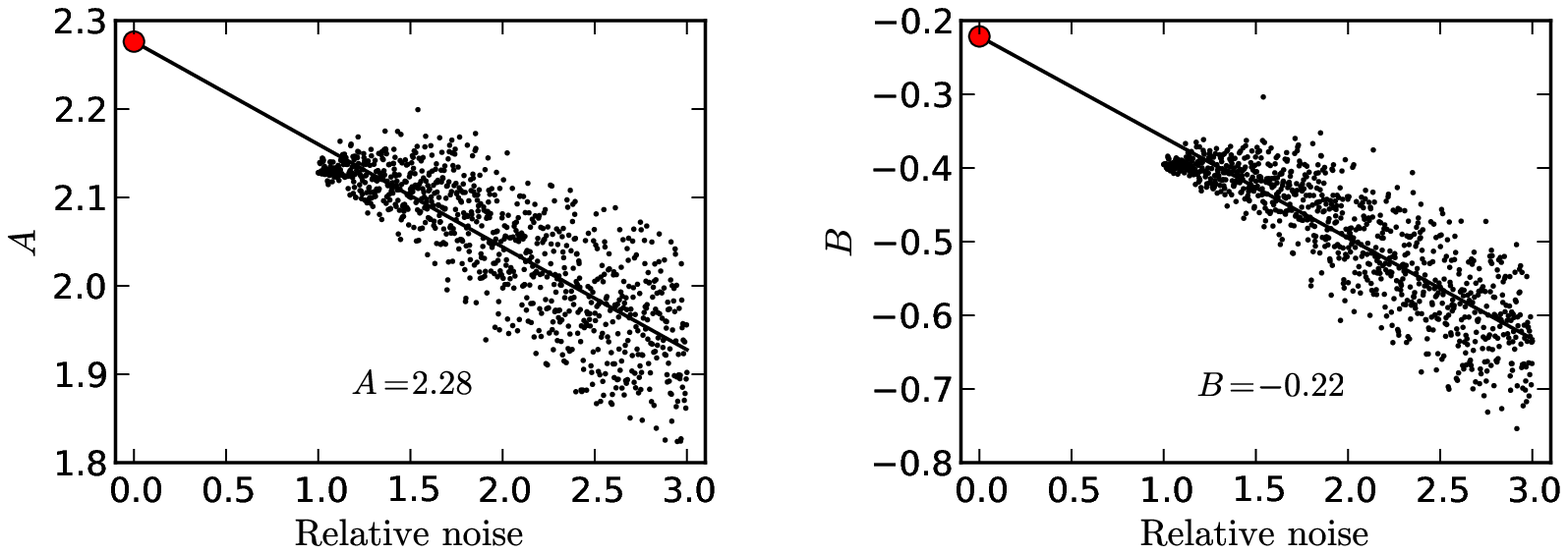}
\caption{
The results of the SIMEX method for the same case as in
Fig.~\ref{fig:chi2}. The points correspond to the $A$ and $B$ values
derived in simulations of varying noise level. The red circles
indicate the final values obtained by an extrapolation to a noise
level of zero.
}
\label{fig:SIMEX}%
\end{figure}

\begin{figure}
\centering
\includegraphics[width=8.7cm]{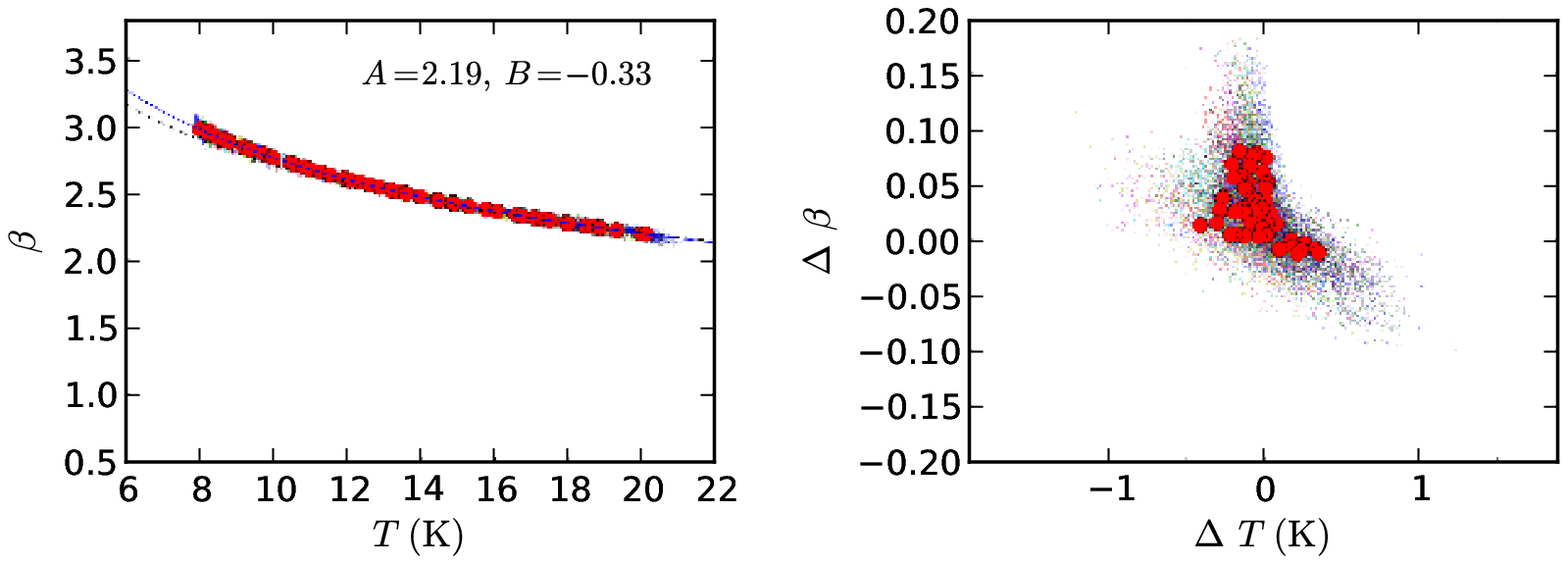}
\caption{
The results of the MC method for the same case as the previous figures.
{\em Left frame:} MC estimates of the $T$ and $\beta$ values. The
red circles show the average parameters for each source. The dots
represent 1000 MCMC samples, the colours corresponding to different
sources. The dashed black line is the true $\beta(T)$ relation and the
solid blue line is the estimated relation. 
{\em Right frame:} Absolute errors in the $T$ and $\beta$ values of
individual sources. The dots correspond to 1000 MCMC estimates for the
parameters of each source and the solid red circles are the average
values. 
}
\label{fig:v2}%
\end{figure}

\begin{figure}
\centering
\includegraphics[width=8.7cm]{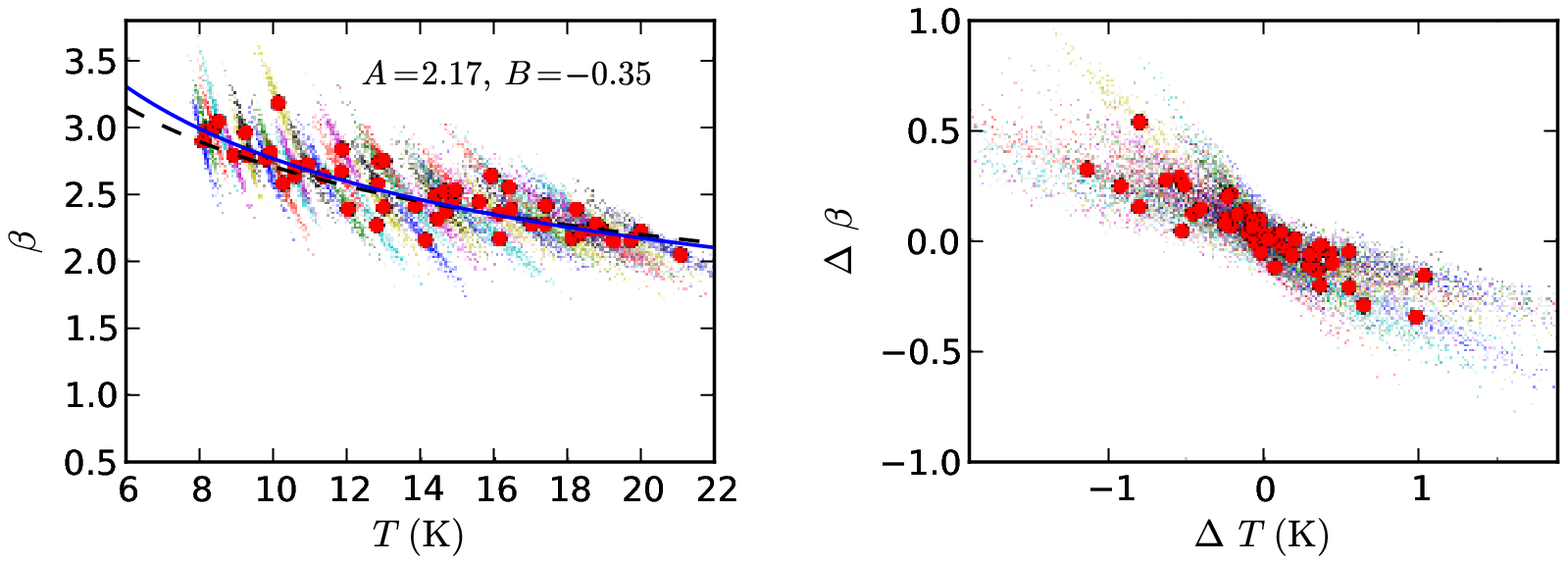}
\caption{
The results of the BM method for the same case as in the previous
figures. The contents of the frames are as in Fig.~\ref{fig:v2}.
}
\label{fig:v3x}%
\end{figure}

\begin{figure}
\centering
\includegraphics[width=8.7cm]{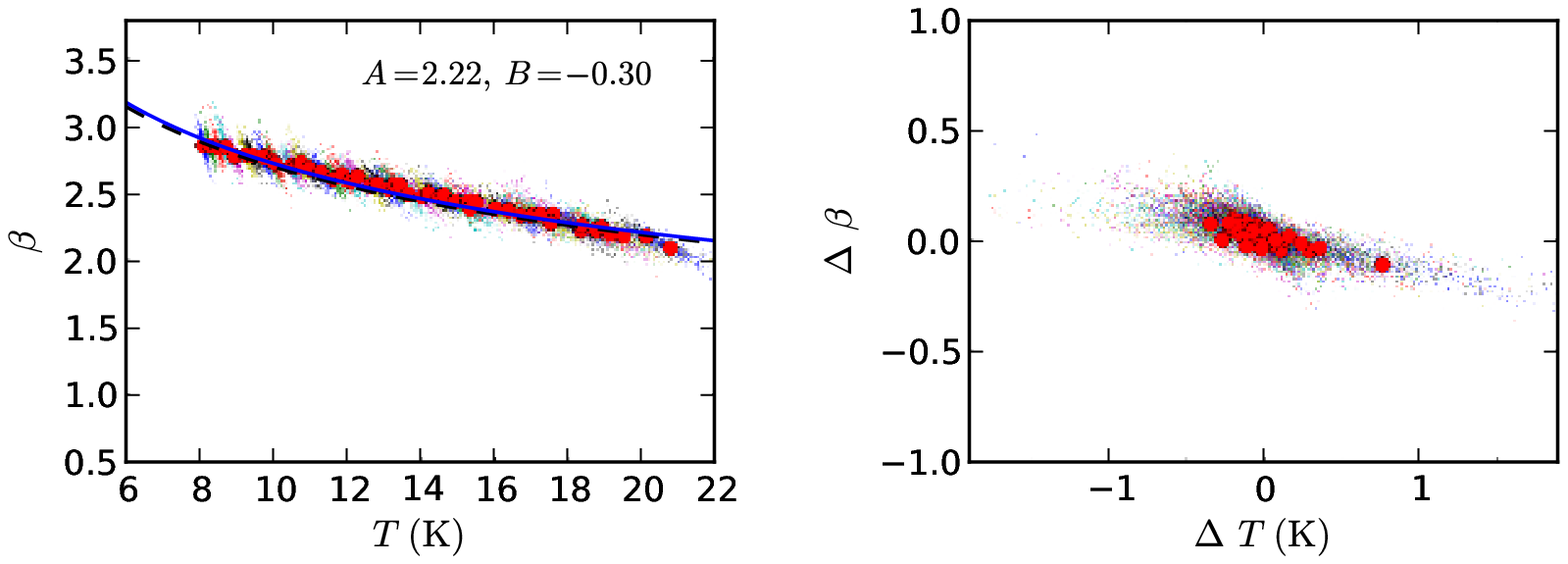}
\caption{
As Fig.~\ref{fig:v3x} but for HM(S), the hierarchical model using
Student probability distributions.
}
\label{fig:v3}%
\end{figure}

Figures~\ref{fig:chi2}--\ref{fig:v3} show examples of the results
obtained with the $\chi^2$, SIMEX, BM, MC, and the HM methods when
observations made at Herschel wavelengths have a constant
signal-to-noise ratio across all bands and sources. The data consist
of measurements of 50 sources with temperatures uniformly distributed
between 8\,K to 20\,K and with the spectral indices following the
$\beta= A (T/20)^{B}$ law with values of $A$=2.2 and $B$=-0.3. 
The relative noise in the measurements is 5\%. The figures
indicate the $\beta(T)$ relations recovered by each of the methods and
the temperature and spectral index errors of individual sources.

The figures \ref{fig:chi2}--\ref{fig:v3} all correspond to the same one realisation
of observational noise. Some differences are already visible between the methods
but we must examine several noise realisations to estimate the bias of the methods.

In the following figures we compare the error distributions in
the cases the different methods. Each synthetic set of observations
consists of data on 50 sources. We derive the parameters $A$ and $B$
for 40 data sets that contain different realisations of observational
noise. The scatter of the obtained parameter values provides crude
estimates of the error distributions of the parameters $A$ and $B$.
Because each individual realisation already consists of observations
of 50 sources, the variation in the estimated parameters remains
small. This gives us confidence that the data set is sufficient to
derive conclusions on the relative performance of the methods.

\begin{figure}
\centering
\includegraphics[width=8.7cm]{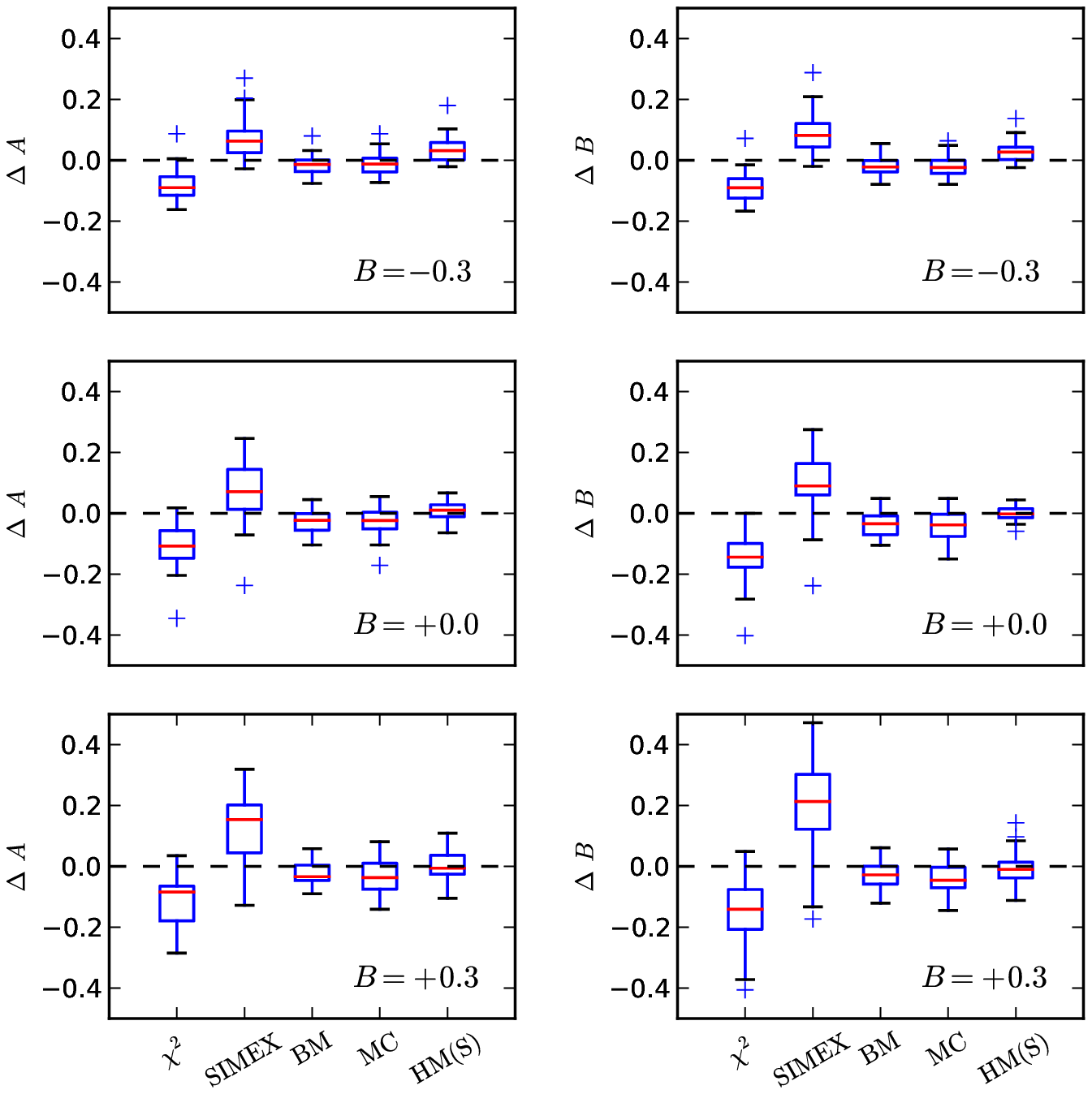}
\caption{
Comparison of the errors in the recovered parameters $A$ (left hand
frames) and $B$ (right hand frames) of the $\beta(T)$ relation. The
observations consist of measurements at 100, 160, 250, 350, and
500\,$\mu$m, with a relative noise of 5\%. The true value of $B$ is
indicated in the frames. 
}
\label{fig:stats_H_N1}%
\end{figure}

\begin{figure}
\centering
\includegraphics[width=8.7cm]{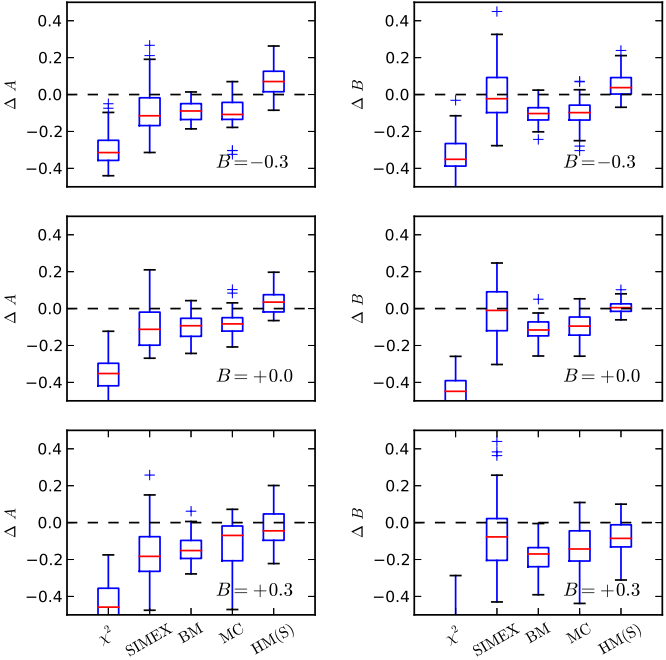}
\caption{
As Fig.~\ref{fig:stats_H_N1} but for a relative noise of 10\%.
}
\label{fig:stats_H_N2}%
\end{figure}

\begin{figure}
\centering
\includegraphics[width=8.7cm]{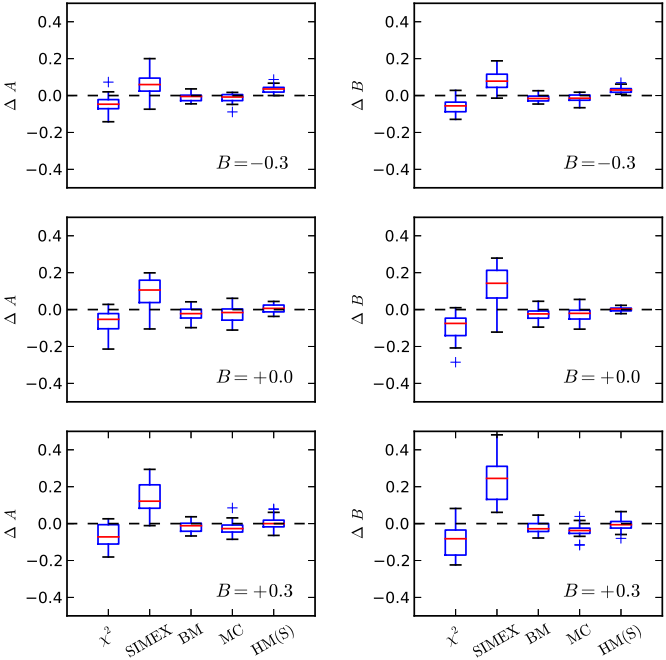}
\caption{
As Fig.~\ref{fig:stats_H_N1} but showing results for observations at
100, 350, 550, and 850\,$\mu$m with a relative noise of 5\%.
}
\label{fig:stats_P_N1}%
\end{figure}

\begin{figure}
\centering
\includegraphics[width=8.7cm]{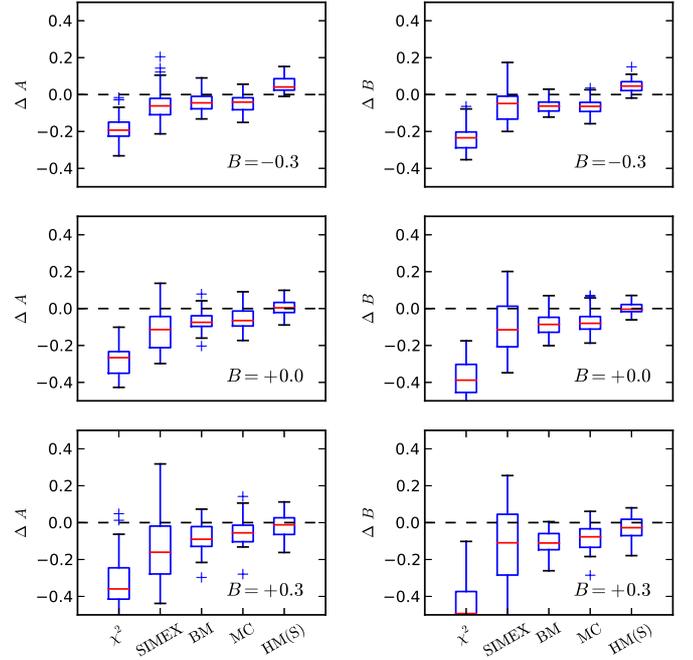}
\caption{
As Fig.~\ref{fig:stats_P_N1} (observations at 100, 350, 550, and
850\,$\mu$m) but showing results for a relative noise of 10\%.
}
\label{fig:stats_P_N2}%
\end{figure}

Figures~\ref{fig:stats_H_N1} and \ref{fig:stats_H_N2} show the results for the
analysis of simulated observations at 100, 160, 250, 350, and 500\,$\mu$m. The
figures correspond, respectively, to cases where the relative uncertainty of the
measurements is either 5\% or 10\%. The plots show the distributions of the $A$ and
$B$ parameter values relative to the true value in the simulations. The boxplots
show the errors in the recovered parameters $A$and $B$ for the methods: $\chi^2$,
SIMEX, MC, BM, and HM(S). The boxplots show the median value of the recovered
parameter (horizontal line inside the boxes), the range of values from the first to
the third quartile (the height of the box), and the outliers (crosses) that are
more than 1.5 times the extent of the inner quartile range (the distance shown as
vertical lines or ``whiskers'') outside the box.
Figures~\ref{fig:stats_P_N1} and \ref{fig:stats_P_N2} show corresponding results for
observations at 100, 350, 550, and 850\,$\mu$m. The relative noise of the measurements is
again either 5\% or 10\%.

In Figs.~\ref{fig:stats_H_N1}--\ref{fig:stats_P_N2} the behaviour of
the methods MC and BM is very similar to each other. The parameters
$A$ and $B$ are both slightly underestimated although this is clear
only in the case of the larger relative error. The bias of the
hierarchical models is of similar magnitude but the sign of the bias
tends to be opposite to the sign of $B$.

\subsection{Analysis of smaller temperature ranges}
\label{sect:Tsplit}

We examined the observations of the previous section also splitting
them to two samples with temperatures above and below 14\,K. The
smaller dynamical range should make it more difficult to reliably
estimate the $\beta(T)$ relation. In particular, we are interested in
the performance of the hierarchical model because it includes a 
global model for the distribution of the ($T$, $\beta$) values.

\begin{figure}
\centering
\includegraphics[width=8.7cm]{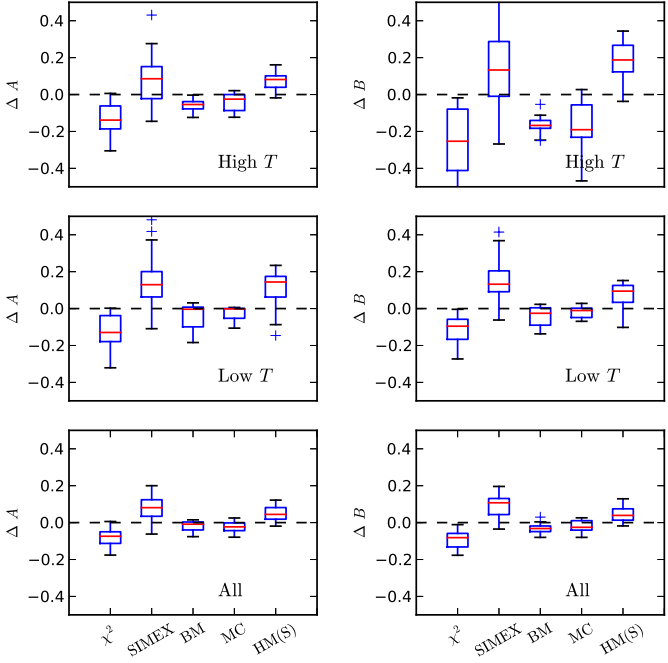}
\caption{
The accuracy of the parameters of the $\beta(T)$ relation for sources
in different temperature ranges. The results are based on measurements
at 100, 160, 250, 350, and 500\,$\mu$m with 5\% relative noise and a
$\beta(T)$ relation with $B$=-0.3. The high and low temperature ranges
correspond to temperatures above and below 14\,K. The bottom frames
show the results for the combined sample.
}
\label{fig:stats_H_N1_hilo}%
\end{figure}

Figure~\ref{fig:stats_H_N1_hilo} shows the errors of the recovered $A$
and $B$ parameters for $\chi^2$, SIMEX, BM, MC, and HM(S) methods. The
full set of observations consists of 50 sources with a fixed relative
uncertainty of 5\% for intensity measurements at 100, 160, 250, 350,
and 500\,$\mu$m, following the $\beta(T)$ law with $B=$-0.3. The
results are shown for the 25 sources with temperatures in the range
8--14\,K, the other 25 sources in the range 14--20\,K, and for the
combined sample of 50 sources. The statistics are based on 30 noise
realisations.
As previously, (e.g., Figs.
~\ref{fig:stats_H_N1}-\ref{fig:stats_P_N2}) in the case of the same
value of $B$, MC and BM show a small negative bias and the HM(S) a
slightly larger positive bias. This applies to both the $A$ and $B$
parameters.

\subsection{Observations with a fixed absolute sensitivity}
\label{sect:anoise}

\begin{figure}
\centering
\includegraphics[width=8.7cm]{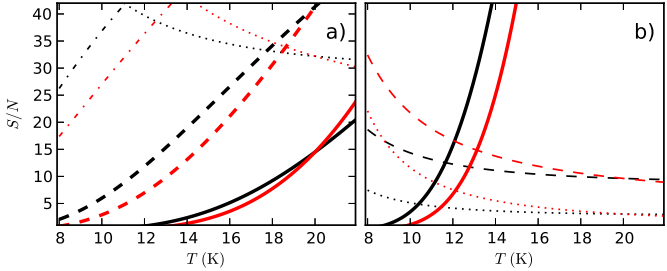}
\caption{
The signal-to-noise ratio of the simulated observations of
Sect.~\ref{sect:anoise} as a function of the source temperature. The
left hand frame corresponds to the simulated {\it Herschel} observations at
100\,$\mu$m (solid line), 160\,$\mu$m (dashed line), 250\,$\mu$m
(dash-dotted line), and 500\,$\mu$m (dotted line). The right hand
frame shows the signal-to-noise ratios for the simulated IRAS
100\,$\mu$m data (solid line) and the {\it Planck} 500\,$\mu$m and
850\,$\mu$m bands (dashed and dotted lines, respectively). The
signal-to-noise ratio of the 350\,$\mu$m observations is 40. The black
lines correspond to the relation $\beta=2.2(T/20.0)^{-0.3}$ and the
red lines to $\beta=2.2(T/20.0)^{+0.3}$.
}
\label{fig:SN}%
\end{figure}

In this section we adopt noise levels that are more in line with 
actual observation with {\it Herschel} and {\it Planck}. In the
case of {\it Herschel}, we start with the sensitivities of 8.1, 3.7, 1.20,
0.85, and 0.35\,MJy/sr at the wavelengths of 100.0, 160.0, 250.0,
350.0, 500.0\,$\mu$m, respectively \citep[as in, e.g.,][]{Malinen2011,
Juvela2012_chi2}. For extended emission, the noise per unit
area would be relatively smaller at the short wavelengths because of
the smaller beam size. However, we use directly the ratio of the
values listed above that is more appropriate for point-like sources
but also accentuates the differences in the signal-to-noise ratios
between the wavelengths. Similarly, for IRAS 100\,$\mu$m and {\it Planck}
data at 350.0, 550.0, and 850\,$\mu$m, we start with uncertainties of
0.06, 0.12, 0.12, 0.08\,MJy\,sr$^{-1}$ \citep[as
in][]{Juvela2012_chi2}. The actual noise levels are fixed by setting
the signal-to-noise ratio to a value of 40 at the wavelength of
350\,$\mu$m. At the other wavelengths, the signal-to-noise ratio
varies depending on the source temperature and the $\beta(T)$ law. In
particular, the signal-to-noise ratio becomes small for the coldest
sources and the shortest wavelengths (see Fig.~\ref{fig:SN}).

\begin{figure}
\centering
\includegraphics[width=8.7cm]{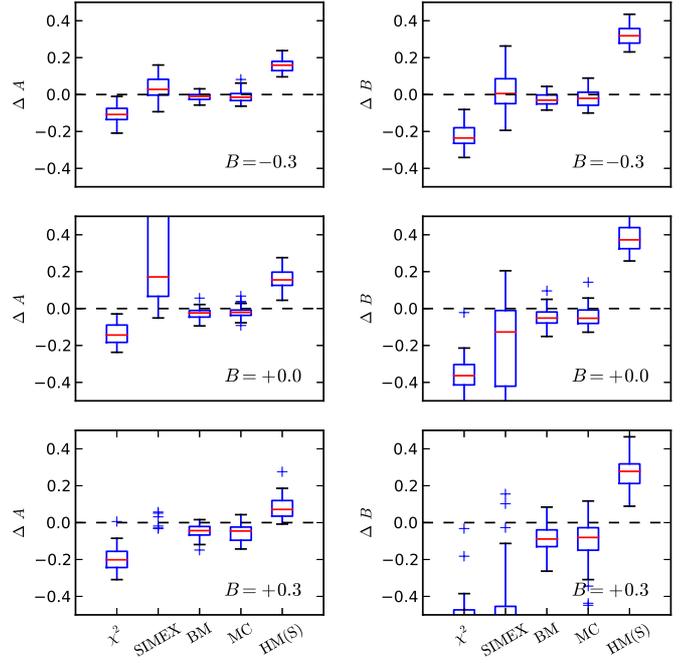}
\caption{
The errors of the parameters of the $\beta(T)$ relation in case of
{\it Herschel} observations with fixed absolute noise levels.
}
\label{fig:H_N4}%
\end{figure}

\begin{figure}
\centering
\includegraphics[width=8.7cm]{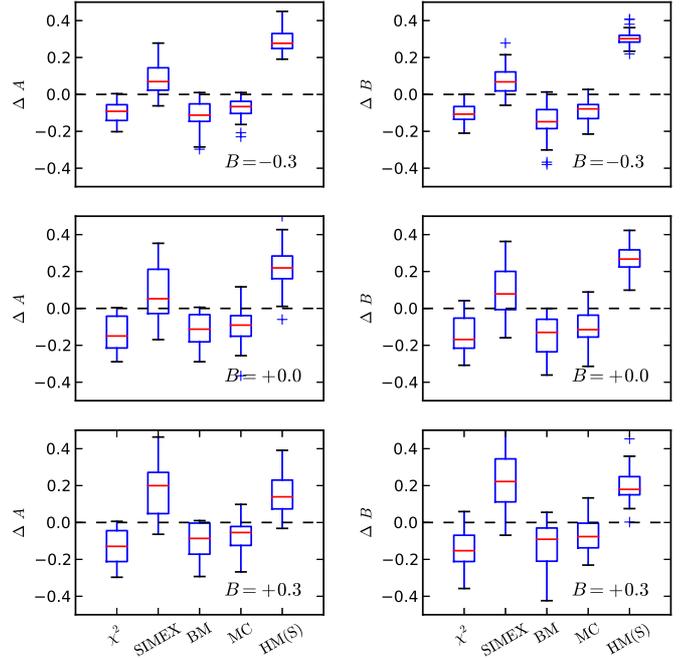}
\caption{
The errors of the parameters of the $\beta(T)$ relation in the case of
IRAS and {\it Planck} observations with fixed absolute noise levels.
}
\label{fig:P_N4}%
\end{figure}

The accuracy of the recovered parameters of the $\beta(T)$ relation 
are shown in Figures~\ref{fig:H_N4} and \ref{fig:P_N4}, for the
{\it Herschel} and the {\it Planck} cases, respectively. The results of the SIMEX
method are often poor and fall partly outside the plotted range. This
appears to be caused by the very low signal-to-noise ratio of some
sources that, when the noise is further increased, leads to outliers
that destroy the accuracy of the extrapolation. The situation might be
improved by using larger set of noise realisations and, in the fits of
$\beta(T)$, replacing the least squares with a more robust fitting
routine. 
For the other methods the results are consistent with the previous
tests with fixed relative measurement uncertainties. BM and MC are
close to each other. The same applies to the pair of HM(N) and HM(S).
However, the hierarchical models show somewhat higher bias and, unlike
in Figs.~\ref{fig:stats_H_N1}-\ref{fig:stats_P_N2}, the bias in the
parameter $B$ is always positive irrespective of the true value of $B$

\subsection{The effect of the number of sources} \label{sect:number}

\begin{figure}
\centering
\includegraphics[width=8.7cm]{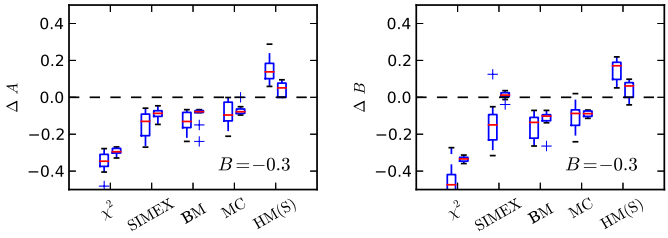}
\caption{
Comparison of the errors in the $A$ and $B$ parameters of the
$\beta(T)$ relation, comparing cases with 50 or 500 observed sources.
The methods of analysis are $\chi^2$, SIMEX, BM, MC and HM(S).  For
each method, two boxplots are shown, the 50-source case on the left
and the 500-source case on the right. The observations consist of data
in {\it Herschel} bands at wavelengths 100--500 $\mu$m and with a
relative noise of 10\%.
}
\label{fig:500sources}%
\end{figure}

In this section, we examine how the number of observed sources (or
pixels) impacts the results. For this purpose, we generated data
consisting of 500 sources with temperatures uniformly distributed
between 8 and 20\,K, and the spectral index following Eq.~\ref{eq:AB}
with $A=2.2$ and $B=-0.3$. Figure~\ref{fig:500sources} compares the
results obtained with methods $\chi^2$, SIMEX, BM, MC and HM(S) when
applied to data in the {\it Herschel} bands at wavelengths 100--500
$\mu$m and with a relative noise of 10\%. Ten realisations are used in
this case, and the results are compared with the 50-source cases with
the same number of realisations.

As expected, the scatter in $A$ and $B$ values is significantly
smaller in the case of 500 sources, because of the averaging of the
statistical noise. More interestingly, all methods also exhibit
smaller bias. This is particularly true for the $\chi^2$ and SIMEX
methods, SIMEX having almost no bias left. The HM(S) method also
performs better, producing results compatible with the true values of
$A$ and $B$. The MC and BM results are more robust regarding the
source numbers, the variation in bias between the 50- and 500-source
cases being compatible with the scatter of results.

\subsection{The effect of the distribution of source temperatures}
\label{sect:NT}

The hierarchical model includes the implicit assumption that the joint
probability distribution of $I$, $T$, and $\beta$ can be described as
a multidimensional normal distribution (version HM(N)) or as Student
distribution (version HM(S)). By using sources whose temperatures are
distributed uniformly between 8\,K and 20\,K, we are clearly
inconsistent with this assumption. To see whether this has an
effect on the results, we examined a case where the source
temperatures are extracted from a normal distribution with a mean value of
14.0\,K and a standard deviation of 3.0\,K, the values being
restricted to the same range 8.0--20.0\,K. 

\begin{figure}
\centering
\includegraphics[width=8.7cm]{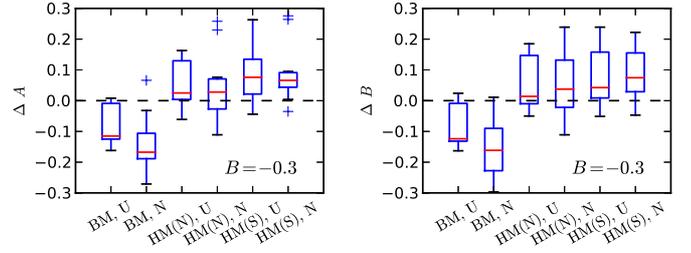}
\caption{
Comparison of the errors in the $A$ and $B$ parameters of the
$\beta(T)$ relation, comparing cases where the source temperatures
follow a uniform (U) or a normal (N) distribution. The compared
methods of analysis are BM, HM(N), and HM(S). The observations
consist of data in {\it Herschel} bands at wavelengths 100--500\,$\mu$m and
with a relative noise of 10\%.
}
\label{fig:NT}%
\end{figure}

Figure~\ref{fig:NT} compares the cases between uniformly and normally
distributed source temperatures. The results are shown for the BM,
HM(N) and HM(S) methods when the relative noise at {\it Herschel}
wavelengths is 10\%.  Each simulated set of observations consists of
50 sources. Figure~\ref{fig:NT} is based on data from 30 noise
realisations. In the case of the hierarchical models, the results do
not indicate any clear dependence on the form of the temperature
distribution. Because the normal distribution also means an
effectively smaller dynamical range of temperatures, it may even
suggest that the loss of dynamical range is compensated by the better
match with the assumptions of the method, i.e., the normal
distribution of temperature values. However, it is clear that the bias
and the dispersion of the parameter estimates are not strongly
affected by the change of the temperature distribution. The only
noticeable effect is the marginally larger bias of the BM results
when the distribution of source temperatures follows normal
distribution.

\begin{figure}
\centering
\includegraphics[width=8.7cm]{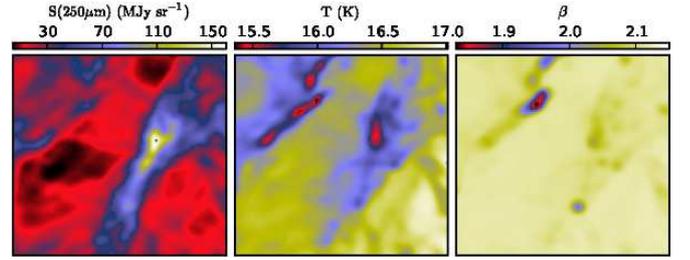}
\caption{
The 250\,$\mu$m surface brightness, dust colour temperature, and spectral
index maps for the MHD model cloud. The values correspond to the $\chi^2$
solution with noiseless data.
}
\label{fig:chi2_N0}%
\end{figure}

\begin{figure*}
\centering
\includegraphics[width=14cm]{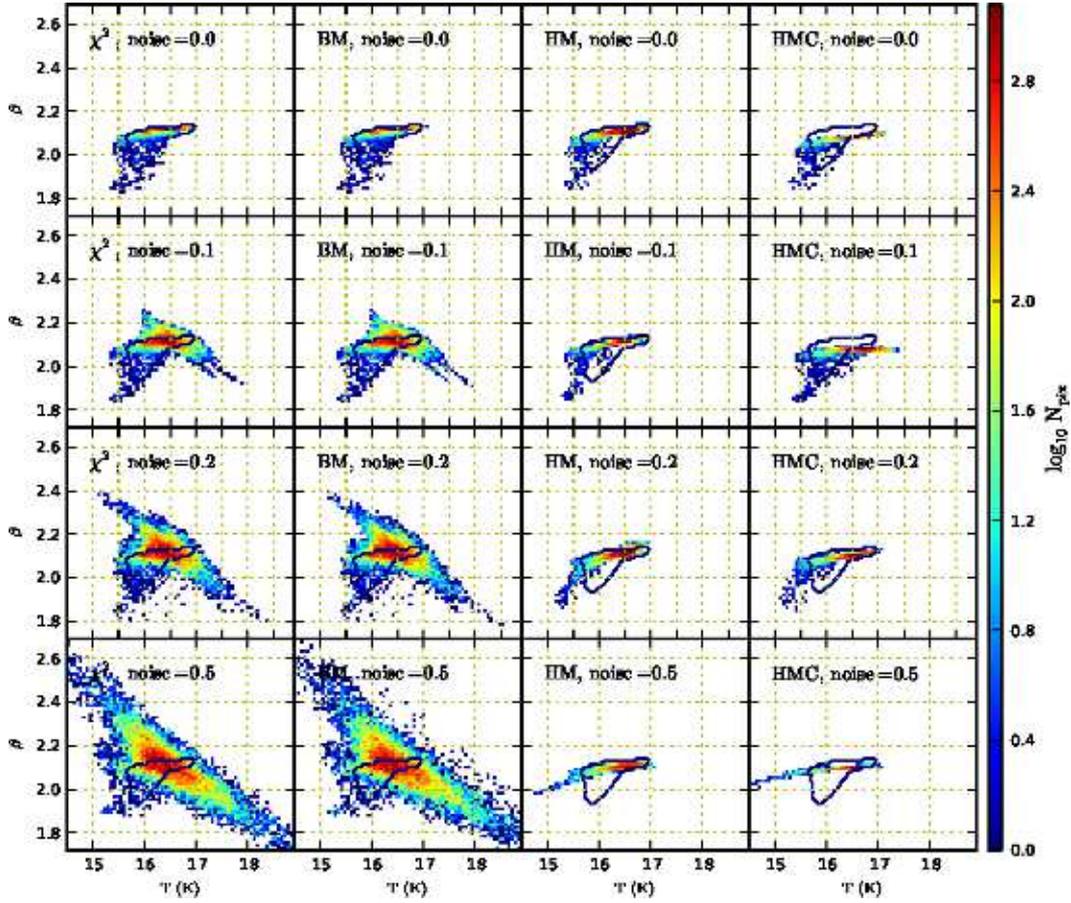}
\caption{
The distributions of ($T$, $\beta$) values for different methods ($\chi^2$, BM,
HM(N), and HMC) and for the MHD model and noise levels $N$=0.0-0.5. The values are
the Bayesian estimates of the individual pixels.To guide the eye, a single
contour following the solution for the $N=0.0$ case is plotted in each frame.
The colour scale corresponds to logarithmic number of pixels per a ($T$,
$\beta$) interval.
}
\label{fig:simu}%
\end{figure*}


\section{Results for a radiative transfer model} \label{sect:results_MHD}

In this section we analyse synthetic surface brightness data that are obtained with
magnetohydrodynamic simulations and radiative transfer modelling (see
Sect.~\ref{sect:MHD_model}). The comparison is mainly restricted to $\chi^2$, BM, and HM(N)
methods. We also examine the effects of mapping artefacts and calibration errors. To this end, we
carry out calculations using HM including free parameters for the relative calibration of the
observed bands. This variation of the HM method is denoted as HMC.

Figure~\ref{fig:chi2_N0} shows the temperature and spectral index maps calculated
with the $\chi^2$ method using noiseless data ($N$=0.0). The colour temperature
ranges from 15.36\,K to 17.01\,K, lower temperatures being found towards regions
of higher column density. The values of the observed spectral index are between
1.82 and 2.13, the low values also being associated with dense regions. In the
model, the intrinsic emissivity spectral index of dust grains themselves was
constant, close to the observed maximum value. In observations, the deviations
from the intrinsic value result from line-of-sight temperature variations that
tend to decrease the apparent spectral index deduced from observed intensities.
In the fit, the different bands were weighted according to the noise values given
above. Although these data contain no noise, the relative weighting may slightly
affect the results because the simulated intensities do not precisely follow a
modified black body law.
Still, we consider these maps as references to which all methods are compared.
The top-left panel of Fig.~\ref{fig:simu} shows the $T-\beta$ distribution of
these maps. Two main features characterise the maps. Firstly, there is a tight
$T-\beta$ correlation for most most pixels, these corresponding to areas of
lower column density and relatively high values of both $T$ and $\beta$. The
second features corresponds to a smaller number of pixels that are found in
regions of high column density and that have lower values of $T$ and especially
of $\beta$.

Figure~\ref{fig:simu} shows the distribution of the recovered ($T$, $\beta$)
values for the $\chi^2$, BM, and HM methods for noise levels $N=$0.0, 0.1,
0.2, and 0.5. When the input data contain no noise, $N=0.0$, the calculations
were completed using error estimates corresponding to the $N=0.1$ case. This
was done because the calculations are technically not possible if error
estimates are zero. The values plotted in Fig.~\ref{fig:simu} are the Bayesian
estimates of each map pixels, i.e., the mean values of the corresponding
parameter values in the MCMC chains.  

Unlike in Sect.~\ref{sect:results_mbb}, the results of $\chi^2$
and BM are almost identical. Section~\ref{sect:results_mbb} was
discussing the relative bias between sources of different temperatures
while data in Fig.~\ref{fig:simu} correspond to a narrow temperature
range (see Fig.~\ref{fig:chi2_N0}). Over an interval of $\Delta T =
1.0$\,K a bias of $\Delta B=0.1$ corresponds to a change of less than
0.01 in the spectral index, too small to be noticeable. Furthermore,
in Fig.~\ref{fig:simu} the noise level is lower, especially relative
to the 10\% relative noise case of Sect.~\ref{sect:results_mbb}.
Asymmetries in the distribution of $\chi^2$ values also are usually
stronger at lower temperatures and, because the temperature and
spectral index values of Fig.~\ref{fig:simu} are well within the range
allowed by the flat priors, the priors are not expected to cause any
differences either.

Unlike BM, HM shows positive correlation between $T$ and $\beta$ up to
the highest noise level. For $N=0.1$, the covariance matrix of the
model implies a correlation coefficient of 0.69 between these
parameters which decreases to 0.52 for $N=0.5$. At both noise levels,
the model covariance matrix gives for $\beta$ a standard deviation of
$\sim$0.015 while pixels with values below $\beta=2$ only exist at the
lower noise level. Therefore, the difference in the distribution of
$\beta$ values is not caused by a significant change in the global
part of the hierarchical model but rather directly by changes in the
relative weight given for the likelihood function.

\begin{figure}
\centering
\includegraphics[width=8.7cm]{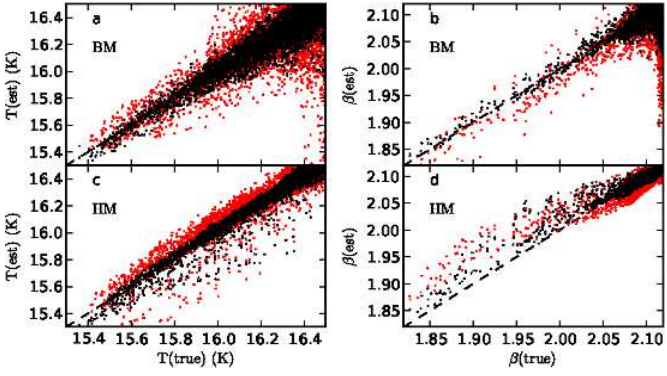}
\caption{
The parameters estimated from noisy data versus the true values obtained from
noiseless observations. The upper frames correspond to the BM method and the
lower frames to the HM method. In each frame, black and red points correspond,
respectively, to noise level of $N=0.1$ and $N=0.2$. 
}
\label{fig:TB_scatter}%
\end{figure}

Figure~\ref{fig:TB_scatter} compares the results of BM and HM(N) in case of noise
levels $N=0.1$ and $N=0.2$. The plots shows the behaviour emphasising regions with
low $T$ and $\beta$ values. One should note that the pixels with $T<16.0$\,K and
pixels with $\beta<2.05$ represent only a small fraction of the full map, $\sim$8\%
and $\sim$2\%, respectively. On the other hand, being associated with dense cloud
structures, the signal-to-noise ratios of those pixels are higher than average.
Figure~\ref{fig:TB_scatter} shows that HM provides significantly lower scatter at
both noise levels, this resulting in more accurate temperature estimates.
However, the HM results also contain bias whereby the spectral index $\beta$ is
overestimated for pixels with low $\beta$. These also are the pixels with the
lowest temperatures.

\begin{figure}
\centering
\includegraphics[width=8.7cm]{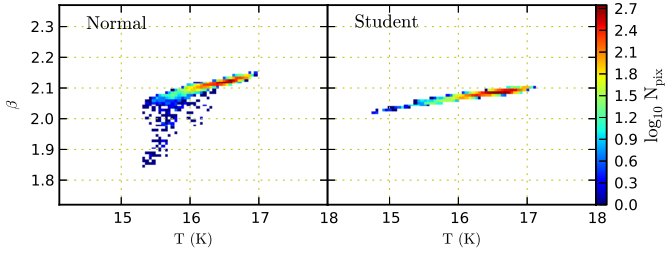}
\caption{
Distribution of recovered $T$and $\beta$ values for MHD model with noise
$N=0.1$ and the HM method with probabilities being calculated either with normal
distribution (left hand frame) or with Student distribution (right hand frame).
The colour scale corresponds to logarithmic number of pixels per ($T$,
$\beta$) interval.
}
\label{fig:MHD_student}
\end{figure}

The results may depend not only on the level of noise but also on the assumption
of the probability distributions that are used in the analysis.
Figure~\ref{fig:MHD_student} shows the distributions of ($T$, $\beta$) values for
the HM(N) method and noise level $N=0.1$. The second frame corresponds to HM(S)
method (i.e., using Eq.~\ref{eq:student1} instead of Eq.~\ref{eq:normal1}). The
use of the Student probabilities results in a solution where all pixels fall onto
an narrow band in the ($T$, $\beta$) plane. The difference was confirmed
repeating the calculations with different initial values. However, for the lowest
temperature pixels, which also correspond to the lowest $\beta$ values, the
convergence of the MCMC chain is slow, requiring several millions of MCMC steps.

\subsection{Effects of noise and calibration errors} \label{sect:MHD_noise}

In Fig.~\ref{fig:simu} we, for the first time in this paper, include results from hierarchical
models that include parameters $\delta$ to account for calibration errors. The
model includes three additional parameters, $\delta_{\rm i}$, the relative scaling
factors of the 160\,$\mu$m, 350\,$\mu$m, and 500\,$\mu$m data. Together these three
factors can compensate for any change in the spectral index, i.e., $\beta$ is 
completely degenerate with these parameters. To obtain a unique solution,
we use for $log_{\rm 10} \delta_{\rm i}$ Gaussian priors with with mean value of
0.0 and standard deviation 0.001.

\begin{figure}
\centering
\includegraphics[width=8.7cm]{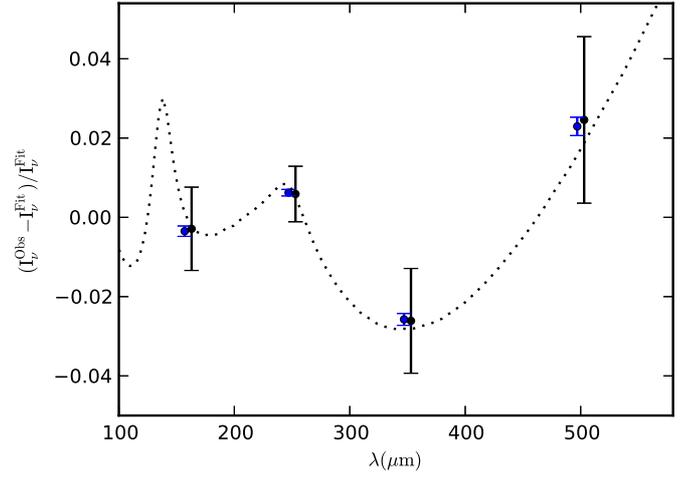}
\caption{
Deviations from the fitted modified black body law in the case of the MHD
model. The plot shows the relative deviations for 5\% of the highest (blue symbols)
and 5\% of the lowest column density (black symbols with larger error bars) pixels.
The error bars correspond to noise per pixel for the general noise level $N=0.1$.
The dotted line shows relative deviations between a $\nu^{2.1}$ dependence and the
actual opacity in the employed dust model.
}
\label{fig:avespe}
\end{figure}

The calibration factors are strongly constrained to remain near the default value
of 1.0 but are, nevertheless, found to deviate from this value even when the input
data contain no actual calibration errors. This is caused by the fact that the
model emission does not follow a modified black body law. The line-of-sight
temperature variations tend to broaden the emission spectrum but here they play
only a minor role. The main cause lies in the opacity of the employed dust model
that does not follow a single $\nu^\beta$ relation \citep{Draine2003}.
Figure~\ref{fig:avespe} demonstrates the deviations of the model spectra relative
to the fitted modified black body spectra. In HMC, these are interpreted as errors
(no longer necessarily calibration errors) relative to the assumed spectral model.
Because each data set consists of more than 10000 pixels, the priors of the
parameters are not strong enough to force the values $log_{\rm 10} \delta_{\rm i}$
to zero.

The results of HMC method do not behave quite consistently as a function of the
noise level. This must be partly related to the changes in the influence of the
$\delta_{\rm i}$ priors for cases of different noise level. However, it also
might indicate some problems in the convergence of the calculations. For some
initial values, the $N=0.5$ calculation appeared to result in a solution that
was $\sim$0.1 below the correct $\beta$ value or gave very nearly the same
$\beta$ value for all pixels. This suggests that the burn-in phase can be very
long and, even after some tens of million MCMC steps, the calculations have not
yet really converged. This is likely to affect mostly noisy observations where
the probability maximum is less well defined. The common feature of all the
solutions was that, as the noise increases, instead of forming a separate group
with clearly lower $\beta$ values, the low temperature pixels begin to conform
with the general distribution dictated by the majority of the pixels.

\begin{figure*}
\centering
\includegraphics[width=11cm]{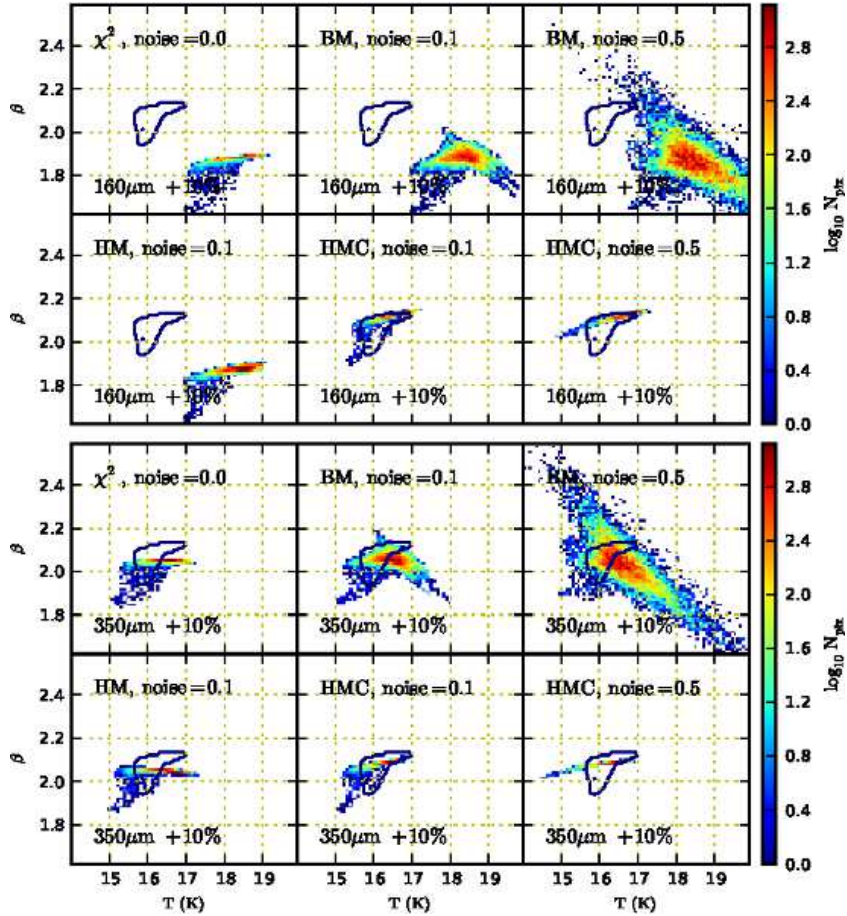}
\caption{
The distributions of ($T$, $\beta$) values in case of data containing calibration
errors where either 160\,$\mu$m data (first six frames) or the 350\,$\mu$m data
(six last frames) has been scaled by 1.1. The methods are $\chi^2$, BM, HM and
HMC, the last one corresponding to hierarchical model that includes free parameters
for multiplicative corrections in three bands. Each frame corresponds to one
combination of method and statistical noise, as indicated in the frames. The single
contour corresponds to the locus of the correct solution (see Fig.~\ref{fig:simu}).
}
\label{fig:simu_cal}%
\end{figure*}

We repeated the HMC calculations for simulations where the intensities of either the
160\,$\mu$m or the 350\,$\mu$m data were multiplied by 1.1. The applied calibration
error is not particularly high but still several times larger than the differences
seen in Fig.~\ref{fig:avespe}. The results are shown in Fig.~\ref{fig:simu_cal}
where, for comparison, we also plot results for selected $\chi^2$ and BM
calculations. HMC is able to recover the ($T$, $\beta$) values quite well,
especially when comparing with the other methods where the effect of the
calibration error is clearly visible. With increased observational noise, the shape
of the ($T$, $\beta$) distribution is distorted in a way similar to the cases in
Fig.~\ref{fig:simu}. Some positive correlation remains between temperature and
spectral index, even at the noise level of $N=0.5$. However, the precise shape of
the distribution was found to depend weakly on the initial values, suggesting
incomplete convergence of the calculations.

\subsection{Effects of mapping artefacts} \label{sect:artefacts}

The MHD model simulations can be considered realistic concerning distribution of
temperature and spectral index values. Because of the temperature variations in the
model, deviations from the simplistic spectral model of a single modified black
body are possible. However, these data are still idealised in the sense that they
do not contain any large scale artefacts that can be encountered in real
observations.

\begin{figure}
\centering
\includegraphics[width=8.5cm]{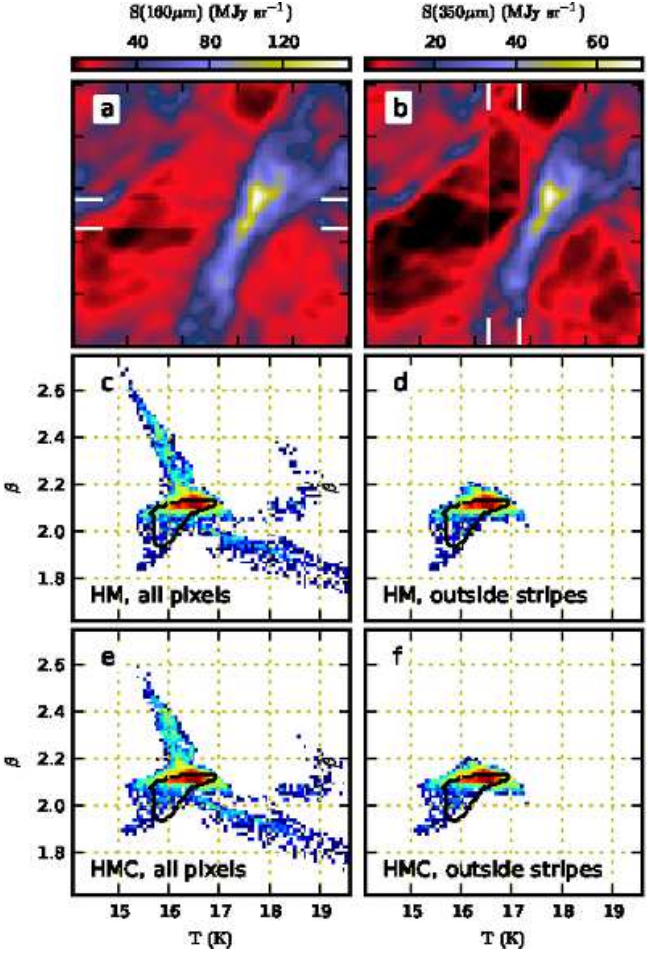}
\caption{
The 160\,$\mu$m (frame $a$) and 350\,$\mu$m (frame $b$) surface brightness maps
($N=0.1$) that include stripes with constant surface brightness offsets. The
white lines indicate the widths of these stripes. Frames $c$ and $d$ show the
distribution of ($T$, $\beta$) values obtained with HM for all the pixels (frame
$c$) and separately for the pixels outside the stripes (frame $d$). Frames $e$
and $f$ show the corresponding results for method HMC. The black contours
correspond to the distribution of values for method HM, in the case of no stripe
artefacts and no noise (see Fig.~\ref{fig:simu}). In frames $c$-$f$, the colour
scale corresponds to logarithmic density of pixels per a ($T$, $\beta$)
interval.
}
\label{fig:stripes}%
\end{figure}

We carried out one set of tests on the effects of mapping artefacts. We introduced
in the 160\,$\mu m$ map a horizontal stripe by adding to the affected pixels a
constant value of 3\,MJy\,sr$^{-1}$ and a similar vertical stripe in the
350\,$\mu$m map by subtracting 2\,MJy\,sr$^{-1}$. The width of both stripes is 10\%
of the map size. As shown in Fig.~\ref{fig:stripes}, the artefacts are not strong
enough to be visually prominent in the surface brightness data. The maps were 
analysed with the BM, HM, and HMC methods, with noise levels $N=0.1$ and $N=0.2$.

The temperature and spectral index estimates are strongly affected within the
modified areas, especially at low column densities where the fractional surface
brightness changes are large. Figure~\ref{fig:stripes} demonstrates that, unless
the pixels affected by the stripes are removed, the general appearance of the
($T$, $\beta$) distribution is dominated by these artefacts. It is important to
see whether the results for the remaining pixels are affected through the global
part of the HM model. According to Fig.~\ref{fig:stripes}, there is only a small
effect. Because of the additional free parameters, the stripes could bias the
results of the HMC method more than the results of the HM method. However, in
Fig.~\ref{fig:stripes}, there is no clear evidence of this either. Of course, if
the area or amplitude of the artefacts is increased, more noticeable global
effects should eventually appear.


\section{Results of the analysis of {\it Herschel} observations} \label{sect:results_Herschel}

As the final test we carry out analysis of {\it Herschel} 160\,$\mu$m, 250\,$\mu$m, 350\,$\mu$m, and
500\,$\mu$m observations of the G109.80+2.70 (see Sect.~\ref{sect:Herschel_data}).
   
\begin{figure}
\centering
\includegraphics[width=8.5cm]{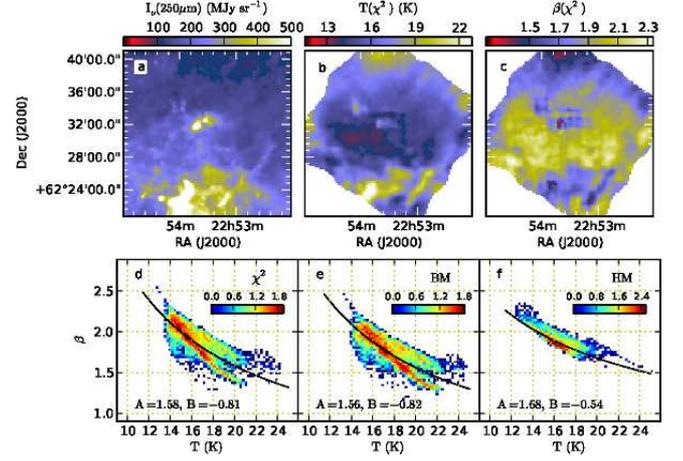}
\caption{
Results for the {\em Herschel} field G109.80+2.70. The upper frames show the
observed 250\,$\mu$m surface brightness and the temperature and spectral
index maps from the $\chi^2$ method. The lower frames show the distributions
of the ($T$, $\beta$) values obtained with the $\chi^2$, BM, and HM(N)
methods, the colour scale corresponding to logarithmic number of pixels per
($T$, $\beta$) interval. The solid lines are least square fits with the
parameters $A$ and $B$ given at the bottom of each frame.
}
\label{fig:PCC288_result}%
\end{figure}

Figure~\ref{fig:PCC288_result} shows the 250\,$\mu$m surface brightness map
(a proxy for column density) and the temperature and spectral index maps
derived with the $\chi^2$ method. The lower frames show the ($T$, $\beta$)
values obtained with the $\chi^2$, BM, and HM(N) methods.
For BM and HM(N) methods the priors are the same as in the previous MHD simulations (see
Table~\ref{table:priors}). Unlike for example in Fig.~\ref{fig:simu}, all methods show a negative
correlation between $T$ and $\beta$. The covariance matrix in the global part of the hierarchical
model gives a value of -0.82 for the correlation coefficient between $T$ and $\beta$. The standard
deviations are 0.058\,K and 0.052 for $T$ and $\beta$, respectively.

We emphasise that the results of Fig.~\ref{fig:PCC288_result} may be affected not only by noise but
also by the mapping artefacts noted in Sect~\ref{sect:Herschel_data}. The observed signal is further
modified by line-of-sight dust temperature variations that decrease $\beta$ especially towards the
embedded sources. Therefore, the observed $\beta(T)$ relation cannot be used directly to infer the
intrinsic opacity spectral index of dust grains. Nevertheless, we also show in
Fig.~\ref{fig:PCC288_result} the results of the least square fitting of Eq.~\ref{eq:AB}. 

\begin{figure}
\centering
\includegraphics[width=8.5cm]{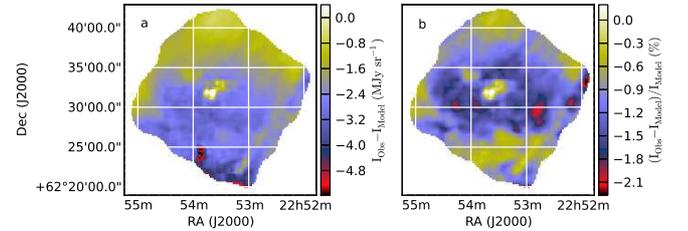}
\caption{
The residuals between the observed 160\,$\mu$m surface brightness in
the field G109.80+2.70 and the $\chi^2$ fit. The residuals are shown
both as absolute errors (left frame) and relative to the $\chi^2$ fit.
}
\label{fig:PCC288_residuals}%
\end{figure}

At 160\,$\mu$m, the differences between the observed surface brightness maps and
the $\chi^2$ fit are shown in Fig.~\ref{fig:PCC288_residuals}. In addition to a
weak large scale gradient, there are some possible artefacts at the map boundaries.
The SPIRE maps do not have any noticeable gradients relative to each other. In the
fit the 160\,$\mu$m map appears to be quite consistent with the 350\,$\mu$m
and 500\,$\mu$m maps while the 250\,$\mu$m map is left with a significant positive
offset. Of course, based on the four maps and without a priori knowledge of the
dust spectral index (or temperature) it is not possible to say whether the
160\,$\mu$m values are in reality too low or the 250\,$\mu$m values too high or if
the SPIRE channels favour a spectral index incompatible with the 160\,$\mu$m data
(e.g., because of the presence of several temperature components). The 160\,$\mu$m
residual map also highlights some smaller features where the relative intensity of
the 160\,$\mu$m is low, either in absolute or relative terms. These may not all be
physically real features but at least the relatively high 160\,$\mu$m residuals
towards the centre of the field can be easily explained by the presence of a warmer
dust component associated with the embedded young stellar objects
\citep{Malinen2011, GCC_II}.

\begin{figure}
\centering
\includegraphics[width=8.5cm]{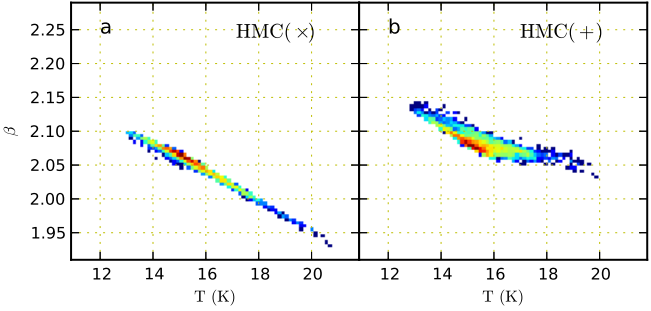}
\caption{
The ($T$, $\beta$) distributions of the field G109.80+2.70 from  hierarchical
models that include multiplicative calibration factors (left frame) or zero point
offsets (right frame) as additional parameters.
}
\label{fig:PCC288_result_3}%
\end{figure}

Surface brightness data may suffer from at least two systematic effects,
multiplicative errors in the surface brightness calibration and additive errors in
the zero point of the surface brightness scale. Assuming that a modified black body
spectrum is an accurate model of the emission, method HMC tries to correct for the
multiplicative errors. However, additive errors may be even a more serious problem
because they introduce systematic effects in temperature and spectral index that
vary as the function of surface brightness, i.e., depending on column density.
Therefore, we also tested a hierarchical model where, instead of the multiplicative
factors, the zero point offsets of the 160\,$\mu$m, 350\,$\mu$m, and 500\,$\mu$m
maps were taken as free parameters. The median surface brightness of these G109.80+2.70
maps were 181, 89, and 40\,MJy\,sr$^{-1}$, respectively. The priors of the offset
corrections were taken to follow normal distribution with the standard deviation
equal to 1\% of the median surface brightness value. The results are shown in
Fig.~\ref{fig:PCC288_result_3}.

In the fits, the additive corrections were +5.7, +6.3, and +1.2\,MJy\,sr$^{-1}$
for the 160\,$\mu$m, 350\,$\mu$m, and 500\,$\mu$m bands, respectively. For the
160\,$\mu$m band this is consistent with Fig.~\ref{fig:PCC288_residuals}, taking
into account that there the residuals are relative to a fit that included the
160\,$\mu$m data as input. The correction is positive also for the other bands,
suggesting that the 250\,$\mu$m input map has some positive offset with respect
to the other bands. Because of the priors of the additive correction factors, the
solution should be well defined. However, this does not eliminate the fact that
the calibration correction and the derived spectral index are partially
degenerate. According to Fig.~\ref{fig:PCC288_residuals}, the relative errors are
not very significant. This also is reflected in the multiplicative calibration
factors that in our calculations were close to unity, 1.01, 1.05, and 0.98 for
160\,$\mu$m, 350\,$\mu$m, and 500\,$\mu$m, respectively.

If data contains additive errors and model includes only multiplicative
corrections, or vice versa, the solution will have errors that, as a function of
the surface brightness level, behave in a complex fashion. The combination of
models with both additive and multiplicative corrections is left for later
studies. However, it is already clear that an increased number of parameters will
not only make the calculations more time-consuming but also make them more
sensitive to the underlying model assumptions. Problems like large scale
gradients and local mapping artefacts may be more significant problems and they
must be dealt with in a different way.

\section{Discussion} \label{sect:discussion}

We have examined five methods that can be used to estimate the shape of the $\beta(T)$
relation based on multi-wavelength dust continuum observations of a set of sources (or
pixels). The $\chi^{2}$ fitting of the intensity values followed by $\chi^2$ fitting of
the resulting ($T$, $\beta$) values, is known to lead to strongly biased results.
Therefore, we examined especially the bias exhibited by the other methods.

\subsection{Analysis of idealised modified black body spectra}

The performance of the methods was first examined using idealised
observations where each spectrum consisted of a single modified black body and
white noise. The results allowed us to to draw some conclusions on the accuracy
and, in particular, of the bias in the results.

\subsubsection{The SIMEX method}

SIMEX is based on the general idea that while we do not have noiseless data available,
we can examine how the results vary when the noise level is increased beyond that of
the actual observations. The method is empirical, and there is no guarantee that the
behaviour at low and high noise levels is similar so that a reliable extrapolation
would be possible. The noise level of the observations must, of course, also be
estimated, so that one knows how far the extrapolation should be extended.  We modelled
the noise dependence of the $A$ and $B$ values with linear functions.  The results were
partly disappointing. When the observations have 5\% relative uncertainty, SIMEX
overestimates the correction with the result that the bias is of the same order as for
the $\chi^2$ method but of opposite sign. The results were better when the
observational noise was raised to 10\%. For the SIMEX method, the higher noise did not
significantly increase the bias while in the $\chi^2$ fits it was higher by a factor of
about 3--4 (as measured by the parameters $A$ and $B$). Figure~\ref{fig:SIMEX} suggests
that the noise dependence is not entirely linear. This is definitely true close to the
original noise because, as long as the added noise is small compared to the original
noise, the bias should not change much. The reliability of the results also will depend
on the range of noise levels examined, in our case 1--3 times the original noise. When
the data included some observations with very low signal-to-noise ratio
(Figs.~\ref{fig:H_N4} and \ref{fig:P_N4}, see Sect.~\ref{sect:anoise}), the SIMEX
results were very poor. The use of SIMEX in the analysis of real observations would
require a more systematic study of the extrapolation functions, the optimal range of
noise values used, and the amount of Monte Carlo samples required. Our implementation
also may have failed partly because of the non-robust nature of the employed
least-squares fits. At lower noise levels, the systematic behaviour of the bias and the
relatively small amplification of statistical errors suggest that some improvements
over the results presented in this paper should be possible.

\subsubsection{Bayesian model and the MC method}

Compared to the $\chi^2$ and SIMEX methods, Markov Chain Monte Carlo (MCMC)
calculations of the basic Bayesian model (BM) were quite successful in eliminating the
bias in the parameters of the $\beta(T)$ relation. Nevertheless, the estimated values
of the parameter $B$ tended to be slightly too low. In other words, the results showed
a small bias in the direction of the noise, towards a larger anticorrelation (or
smaller positive correlation) between the temperature and the spectral index. The
behaviour was similar in the case of observations with fixed relative noise as in the
tests with fixed absolute noise. The values of $T$ and $\beta$ obtained for individual
sources can show quite significant scatter (e.g., Fig.~\ref{fig:v3x}). However, for a
sample of 50 sources, the statistical uncertainty of the $B$ parameter was below
$\sim$0.1 units for all the examined noise cases. In the method MC, we explicitly
enforce a single $\beta(T)$ relation that, in this paper, also was assumed to be of the
correct functional form, $\beta=A(T/20)^B$. If the assumption is true, this should
result in a more powerful method to recover the values of the parameters $A$ and $B$.
The effect of the noise is taken into account with separate Monte Carlo simulations,
assuming that this can be reliably estimated.

Compared to the hierarchical models, this method has the advantage that we do not need
to specify any model for the noise distribution of $I$, $T$, and $\beta$ values. We
simply use the error estimates of the intensity measurements and use Monte Carlo
simulation to estimate the effect of the noise on the end result, i.e., the parameters
of the $\beta(T)$ relation. This could be useful if the noise induced dispersion of the
$T$ and $\beta$ values cannot be well described with a covariance matrix. For low
signal-to-noise ratios, the joint error regions are known to be strongly curved, the
distributions of individual parameters are skew, and the ($T$, $\beta$) values
themselves are biased. All these effects would be automatically taken into account,
based on the error distributions of the intensity measurements.
In practice, the performance of this MC method turned out to be very similar to that of
BM. In particular, the method tended to underestimate the values of $A$ and $B$ but the
bias was typically no more than $\Delta B=$-0.1 in the case of 10\% relative noise and
less than $\Delta B=$-0.05 for the case with 5\% of relative noise. In spite of the
strong constraint regarding the functional form of $\beta(T)$, the method did not
clearly outperform the other methods. In the absolute noise case, Figs.~\ref{fig:H_N4}
and \ref{fig:P_N4}, the MC method produced results quite comparable to BM but with
smaller scatter in the $T$ and $\beta$ values. The same applies to the tests with 5\%
and 10\% relative noise although there also the hierarchical models performed rather
well. On the other hand, the method MC is relatively slow because each step includes
the fitting of modified black body curves for each source. In this method, the MCMC
chains also were more likely to get stuck, i.e., having a very low rate of acceptance
for the steps taken. This is directly related to the additional condition that forces
the simulated $A$ and $B$ parameters to approach the observed values (see
Sect.~\ref{sect:MC}. The situation might be improved by adjusting the relative step
size of the different Markov chain parameters. The method gives results with low
statistical noise but, on the other hand, the scatter would remain small even if the
sources did not follow the same $\beta(T)$ law.

\subsubsection{Hierarchical models}

The remaining methods were based on the hierarchical model, using two alternative
probability distributions. When the two version were compared (e.g.,
Figs~\ref{fig:H_N4}-\ref{fig:NT}), the results were very similar. This is to be
expected because of the idealised nature of our simulations and, in particular, the
absence of outliers. In the tests with 5\% and 10\% relative observational noise, the
results of the HM method appeared to be slightly biased towards a flat relation. In
other words, the bias was negative when $B$ was positive, and with a positive true
value of $B$, the bias was positive. In the case of 5\% relative noise, both $A$ and
$B$ were recovered with typical bias below 0.05 units. For 10\% of relative noise, the
results were mostly correct to within 0.1 units. However, in the case of simulations
with fixed absolute noise, the bias was increased. This is likely to be caused by the
fact that the signal-to-noise ratio is strongly decreasing towards the lower
temperatures. 
The $(T,\beta)$ relation is then mainly determined by the sources with a high
signal-to-noise ratio, and the low-temperature part is extrapolated based on the
assumed model for the joint $(T,\beta)$ distribution.  The same phenomenon was noted by
\citet{Kelly2012} (see their Fig. 9). 
In this case, the hierarchical models show clear positive bias (see Fig.~\ref{fig:H_N4}
and Fig.~\ref{fig:P_N4}).  Unlike in the previous cases with fixed relative noise, the
sign of the bias does no longer depend on the sign of $B$. The bias is similar for both
the {\it Herschel} and {\it Planck} wavelength sets. For $B$=-0.3 the bias is $\Delta
B=+0.3$ and thus, instead of recovering the decreasing $\beta(T)$ function, the results
would suggest a completely flat relation. This is in stark contrast with the BM and MC
methods which both slightly overestimate the steepness of the $\beta(T)$ relation but
still recover the correct trend, i.e., the sign of $B$.

\begin{figure}
\centering
\includegraphics[width=8.7cm]{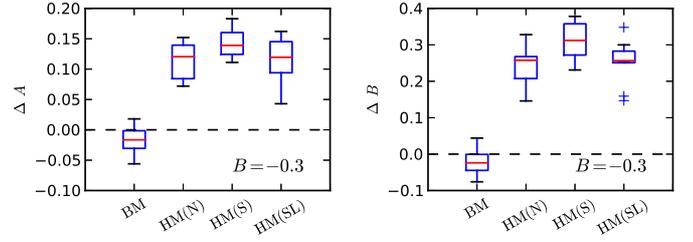}
\caption{
Comparison of results of different versions of the hierarchical
methods, HM(N), HM(S), and HM(SL) where the last one is fitting the
intensity and temperature on logarithmic rather than at linear scale.
The observations consist of {\it Herschel} wavelengths with the absolute
errors as described in Sect.~\ref{sect:anoise}.
}
\label{fig:stats_H_N4_extra}%
\end{figure}

Depending on the version, the hierarchical model tries to describe the distribution of
temperature and spectral index values using either multivariate normal or Student
distribution. \cite{Kelly2012} also examined cases where, contrary to this assumption,
the distribution was bimodal. The method was not found to be sensitive to this
deviation. We carried out a similar comparison between a uniform temperature
distribution and a normal distribution and did not see any clear difference neither in
the bias or scatter of the temperature and spectral index values (see
Fig.~\ref{fig:NT}). In this paper, the default assumption was that of a uniform
temperature distribution. Another difference to the \citet{Kelly2012} calculations is
the fact that while we have used directly the intensity and temperature values, they
used in the fitting the logarithms of these variables.  We tested the difference
between the use of linear and logarithmic scales. For the cases with 5\% and 10\%
relative noise, we do not see any difference in the results obtained with linear and
with logarithmic variables. In the case of absolute (and, for some sources and
wavelengths, larger) errors, there is a small difference (see
Fig.~\ref{fig:stats_H_N4_extra}). However, these differences, if real, are still small
compared to the differences to, e.g., the BM and the true values.
Note that while the ($T$, $\beta$) values follow a non-linear relation, the same also
applies to ($lg\,T$, $\beta$).

\subsection{Distributions of  $T$ and $\beta$ values}
\label{sect:TBdistributions}

\begin{figure*}
\centering
\includegraphics[width=17cm]{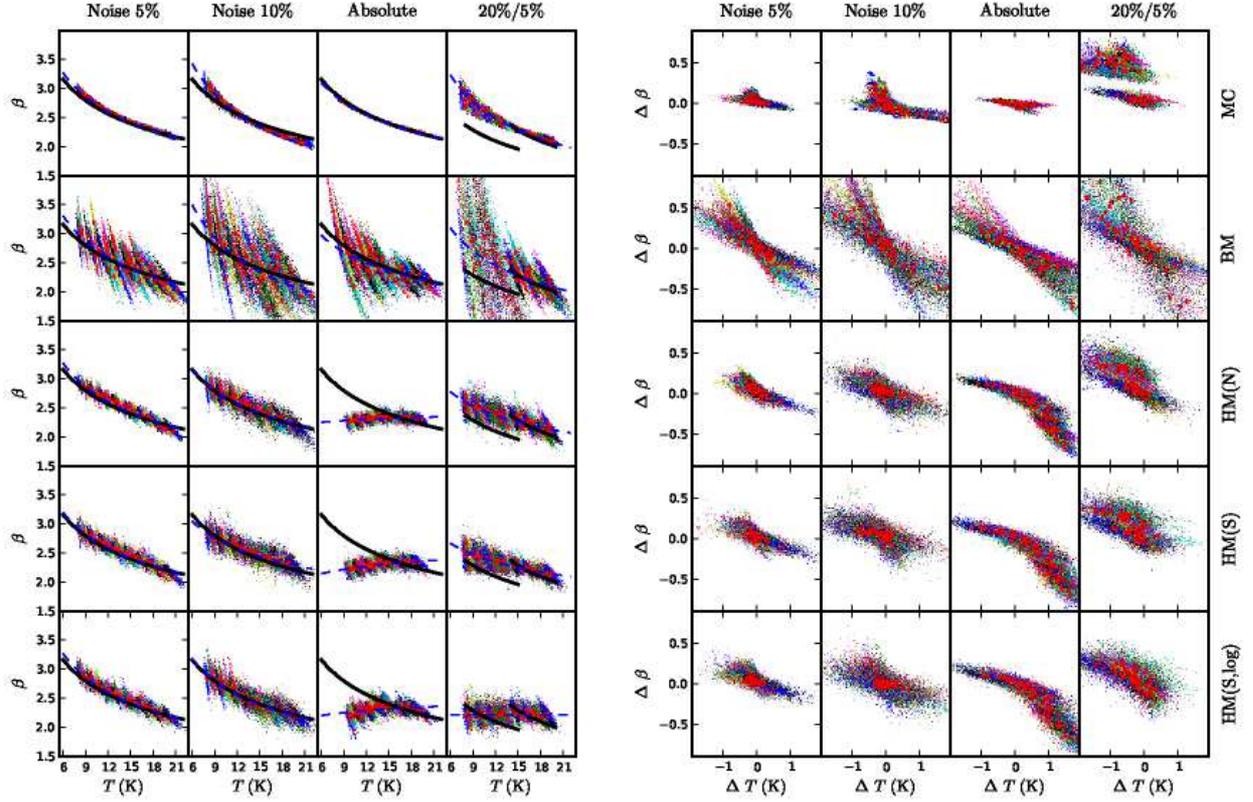}
\caption{
Comparison of the derived ($T$, $\beta$) values in case of
simulated modified black body spectra of different noise level (cf
Sect.~\ref{sect:results_mbb}). The black solid lines correspond to the
relation assumed in the simulation of the observations. The
measurements have 5\% or 10\% relative errors or the absolute errors
described in Sect.~\ref{sect:anoise}. In the fourth column of frames,
the observations follow two distinct $\beta(T)$ relations (black solid
lines) with 20\% relative errors for the low temperature branch
($T<15$\,K) and 5\% relative errors for the high temperature branch
($T>14$\,K). The left hand frames show some MCMC samples (coloured
dots), their averages (red circles) and the fitted $\beta(T)$
relations (dashed blue lines). The right hand frames show the
distributions for the errors $\Delta T$ and $\Delta \beta$. The plots
correspond to a single noise realisation of observations of 50
sources.
}
\label{fig:scatter}%
\end{figure*}

In connection with the simulated modified black bodies, we concentrated mainly
on the bias of the $T(\beta)$ relation. However, the scatter of the individual
($T$, $\beta$) values is equally important. In Fig.~\ref{fig:scatter} we plots
examples of this scatter in the case of MC and BM methods and three variations
of the hierarchical models. The hierarchical models use either normal or
Student probabilities and, in the latter case, the fitted parameters are
either intensity and temperature or their logarithms. This last alternative is
marked as HM(SL).  The figure shows the results for one realisation of the
observations of 50 sources. The columns of the figure correspond to different
noise cases. In the fourth column of Fig.~\ref{fig:scatter}, the 25 sources
below 15\,K also follow a different $\beta(T)$ law from the 25 sources above
14\,K. The parameters are $A=1.8$ and $B=-0.3$ but $\beta$ for the higher
temperature branch is calculated using variable $T-6$\,K instead of $T$. The
relative noise is 20\% for the low temperature and 5\% for the high
temperature branch. 

BM exhibits the largest scatter of individual ($T$, $\beta$) values but the
average estimates are not strongly biased. This applies even to the case of
two $\beta(T)$ relations (fourth columns) because the fits of the sources are
independent from each other. 
The hierarchical models exhibit a significantly
smaller scatter but the results are occasionally more biased. The bias is
usually towards a flat $\beta(T)$ relation. Furthermore, when the data include
some observations that have a higher noise and follow a relation different
from that of the high S/N data, the apparent result is a single $\beta(T)$
relation that also exhibits a relatively small scatter. In the case of real
observations, the sources are unlikely to follow a single $\beta(T)$ law and
one must be cautious in interpreting the results as a proof of such behaviour.
The MC method (uppermost frames in Fig.~\ref{fig:scatter}) is the extreme case
where, by construction, the results will fall on a single $\beta(T)$ relation
whose functional form is predefined. Thus, the results will show a single
relation irrespective of the validity of the assumption.

For hierarchical models, the variation of the observational noise is not
reflected in the scatter of the ($T$, $\beta$) estimates as strongly as in the
case of BM. This is particularly noticeable in the absolute noise case and in
the case of the two $\beta(T)$ relations where the S/N ratio is lower for the
colder sources. The information of the high S/N sources is extrapolated
effectively to the low S/N sources. Depending on the case, this may cause
significant errors. In particular, in the case of two $\beta(T)$ relations of
Fig.~\ref{fig:scatter}, all points incorrectly appear to fall onto a single
$\beta(T)$ relation. This suggests some caution in the interpretation of such
results.

\subsection{Results of the radiative transfer models}

In the MHD models of Sect.~\ref{sect:MHD_model}, the range of temperatures was much
smaller than in the previous tests. Instead of trying to estimate a functional
form of the $T(\beta)$ relation, we only examined the general distribution of the
$(T, \beta)$ values. Compared to the constant 5\% or 10\% relative uncertainties
of Sect.~\ref{sect:rnoise}, the signal-to-noise ratio of the simulated surface
brightness maps varied by a factor of $\sim$30 across the map (see
Fig.~\ref{fig:chi2_N0}), the median being $\sim$13 for 250\,$\mu$m and $N=1.0$.
For the brightest regions, the signal-to-noise ratio exceeds 100 and
these regions strongly influence the overall solution of the hierarchical models.

The BM calculations included flat priors with very wide ranges of allowed
parameter values (see Table~\ref{table:priors}). As a consequence, little
difference was seen between the $\chi^2$ and BM solutions, even for the tails of
the distribution that correspond to data with low signal-to-noise ratios (see
Fig.~\ref{fig:simu}).
The relatively high average temperature also contributes to this result because
the noise-induced error regions tend to become strongly asymmetric only at lower
temperatures. One should remember that the bias discussed in
Sect.~\ref{sect:results_mbb} concerns the relative errors at different parts of the
$T(\beta)$ relation while here, in the absence of noise, all pixels fall in a
temperature interval that has a width not much more than 1\,K.

For the hierarchical models the results are consistent with those discussed in
Sect.~\ref{sect:TBdistributions}. The global part of the hierarchical models helps
to suppress the noise-induced anticorrelation between temperature and spectral
index. In Fig.~\ref{fig:simu}, HM and HMC methods give results where, even when the
noise increases, all pixels fall close to the correct solution.  In the $\chi^2$
and BM solutions the same applies to the highest signal-to-noise pixels but the
figure emphasises the overall scatter that is dominated by the low intensity
regions. In the hierarchical models the global part of the model forces those
regions to follow the distribution of the high signal-to-noise data. The result is
a very tight distribution which is good solution, provided that all pixels behave
statistically in a similar fashion. However, in the case of the MHD model this is
not quite true. The coldest regions also have the strongest line-of-sight
temperature variations and, as a consequence, they deviate from the general
$T-\beta$ -trend \citep[see also][]{Shetty2009a, Juvela2012_TB}. In
Fig.~\ref{fig:simu} these regions form a separate cloud of points below
$\beta=2.0$. The hierarchical models perform very well up to noise level $N\sim
0.2$. When noise is further increased, the low-$\beta$ points are forced to follow
the general trend, i.e., their spectral indices are overestimated and 
temperatures underestimated. In the BM solution, the ($T$, $\beta$) values of
these pixels appear to be more unbiased but the overall distribution is strongly
deformed by the noise already at $N=0.2$.

When the data were modified to include calibration errors, all temperature and
spectral index values were affected. At 160\,$\mu$m a 10\% calibration error
corresponded to a shift of $\Delta T \sim$2\,K and $\Delta \beta \sim 0.2$ (see
Fig.~\ref{fig:simu_cal}). A similar fractional error at 350\,$\mu$m had almost no
effect on temperature and also changed the spectral index estimates only by
$\sim$0.1. When the hierarchical model included parameters for possible
calibration errors, the method was able to recover the correct ($T$, $\beta$)
values almost as well as before the calibration errors were included. However, at
higher noise levels, also the bias in the low-$\beta$ pixels was as noticeable as
in Fig.~\ref{fig:simu}.

The experiment with mapping artefacts demonstrated that, even when the artefacts
are barely noticeable in surface brightness data, they can easily dominate the
appearance of the overall ($T$, $\beta$) distribution (see
Fig.~\ref{fig:stripes}). In our test, the stripes did not significantly affect
the results outside the artefacts themselves. However, if it is possible to
identify artefacts in the input maps (e.g., based on fit residuals) it is always
safest to discard them from the analysis with hierarchical models.

\subsection{Analysis of $Herschel$ maps of G109.80+2.70}

The results for the {\it Herschel} field G109.80+2.70 were presented in
Sect.~\ref{sect:results_Herschel}. In this case the true solution (i.e., in the absence
of noise and possible calibration errors and mapping artefacts) is not known.
Nominally the average signal-to-noise ratio of these data is even higher than in
the case of the MHD models. In the brightest regions the 250\,$\mu$m surface
brightness exceeds 400\,MJy\,sr$^{-1}$ while the noise (estimated at low surface
brightness) is lower by a factor of $\sim$100. On the other hand, the data
appeared to contain some gradients and local artefacts, i.e., errors that cannot
be described easily with any global parameters.

For $\chi^2$ and BM methods, the dispersion of the ($T$, $\beta$) values was
again rather similar although this time the locus of the BM solution corresponded
to slightly higher temperature and lower spectral index (see
Fig.~\ref{fig:PCC288_result}). The difference should correspond to the difference
between the maximum likelihood estimates and the Bayesian estimates that are
averages over MCMC samples. Because the prior distributions were flat over a wide
range of parameter values, the priors should not have a noticeable effect in
Fig.~\ref{fig:PCC288_result}. The Bayesian MCMC estimates often are considered
more robust. However, there also are situations where the maximum likelihood
estimates can be more accurate. One example was shown in \citet{Juvela2012_chi2}
(see Fig. 2 in that article) where the noise induced in the probability
distribution a long tail to lower temperatures and higher values of the spectral
index. In that case the maximum likelihood was reached close to the correct
values while the mean of the probability distribution would be biased. In G109.80+2.70,
the hierarchical model again produced the smallest scatter in the ($T$, $\beta$)
values, although the difference between the methods was smaller than in the
previous MHD models.

We considered alternative hierarchical models with either additive or
multiplicative correction factors for the calibration. Both resulted in similar
temperature values but the derived spectral indices were higher by more than 0.05
when additive corrections were used. The results were positive at least in the
sense that the inclusion of the calibration factors did not lead to widely
different results. However, without strong priors on the calibration factors the
solution could result in very different values of spectral index or even
divergence. One cannot unreservedly advocate the inclusion of calibration factors
as part of routine analysis. The correction is based on the assumption that the
observed radiation follows a modified black body law. This is known to be only an
approximation (because of more complex wavelength dependence of dust opacity and
temperature variations within the telescope beam). The derived calibration
factors also may be affected by possible mapping artefacts, i.e., local features
that cannot be properly taken into account by global parameters. 
It is always useful to examine the residuals between the surface brightness data
and the model fits, whether calibration errors are included in the model or not.
In hierarchical models, it is better to mask the problematic regions rather than
take the risk that the whole solution is affected.

All the methods applied to the G109.80+2.70 data agreed on a general anticorrelation
between $T$ and $\beta$ parameters. The simulation of Fig.~\ref{fig:stripes} acts
as a warning example how strongly the apparent relation between $T$ and $\beta$
can be affected by small imperfections in the data. Nevertheless, we are
confident that in G109.80+2.70 the anticorrelation is real. This is at least partly
connected with temperature variations caused by young stellar objects embedded in
a cold clump (see~\citet{GCC_II} for a toy model of the field). The
correlations between $T$ and $\beta$ in G109.80+2.70 and other fields observed in the
project {\em Galactic Cold Cores} \citep{GCC_III} will be analysed in
more detail in future papers.


\section{Conclusions} \label{sect:conclusions}

We have compared methods that can be used to estimate the $\beta(T)$
relation from a set of dust emission measurements.  These included
direct $\chi^2$ fitting, SIMEX method, Bayesian model, hierarchical
models, and one method (MC) with an explicit assumption of the shape
of the $\beta(T)$ relation. The analysis of simulated observations has
lead to the following conclusions.
\begin{itemize}
\item All the examined methods show some level of bias.  
The sign and the magnitude of the errors depend on the method, the
noise, and shape of the true $\beta(T)$ relation.
\item The methods differ regarding the scatter in the derived
$T$ and $\beta$ values. The hierarchical models produce estimates with
relatively small dispersion. The MC method explicitly enforces a
single $\beta(T)$ law. 
\item 
The hierarchical models are often biased towards a flat $\beta(T)$ relation. When
part of the sources have a low S/N ratio and follow a $\beta(T)$ relation different
from the high S/N sources, the results still show a single relation that is close
to that defined by the high signal-to-noise sources.
\item
Compared to hierarchical models, the Bayesian model typically showed a lower
bias but also a much larger scatter. However, this scatter is more consistent with
the actual uncertainty of the ($T$, $\beta$) estimates of the individual sources.
\item One must keep in mind the special characteristics of the
methods when interpreting their results. It also is always useful to examine the 
residuals of the performed fits.
\item Actual observations may contain both additive and multiplicative
errors and artefacts that affect only parts of the data. As far as possible, these
should be corrected beforehand. In the case of hierarchical models, problematic
data should be masked to prevent those from affecting all results.
\end{itemize}

\begin{acknowledgements}
MJ and JM acknowledge the support of the Academy of Finland grant No.
250741. N. Ysard acknowledges a post-doctoral position from the CNES.
T.L. acknowledges the support of the Academy of Finland grant No.
132291.
We thank the anonymous referee whose comments greatly improved the paper.
\end{acknowledgements}

\bibliography{biblio_v2.0}

\begin{thebibliography}{38}
\expandafter\ifx\csname natexlab\endcsname\relax\def\natexlab#1{#1}\fi

\bibitem[{{Andr{\'e}} {et~al.}(2010){Andr{\'e}}, {Men'shchikov}, {Bontemps},
  {K{\"o}nyves}, {Motte}, {Schneider}, {Didelon}, {Minier}, {Saraceno},
  {Ward-Thompson}, {di Francesco}, {White}, {Molinari}, {Testi}, {Abergel},
  {Griffin}, {Henning}, {Royer}, {Mer{\'{\i}}n}, {Vavrek}, {Attard},
  {Arzoumanian}, {Wilson}, {Ade}, {Aussel}, {Baluteau}, {Benedettini},
  {Bernard}, {Blommaert}, {Cambr{\'e}sy}, {Cox}, {di Giorgio}, {Hargrave},
  {Hennemann}, {Huang}, {Kirk}, {Krause}, {Launhardt}, {Leeks}, {Le Pennec},
  {Li}, {Martin}, {Maury}, {Olofsson}, {Omont}, {Peretto}, {Pezzuto}, {Prusti},
  {Roussel}, {Russeil}, {Sauvage}, {Sibthorpe}, {Sicilia-Aguilar}, {Spinoglio},
  {Waelkens}, {Woodcraft}, \& {Zavagno}}]{Andre2010}
{Andr{\'e}}, P., {Men'shchikov}, A., {Bontemps}, S., {et~al.} 2010, \aap, 518,
  L102

\bibitem[{{Andr\'e} {et~al.}(2000){Andr\'e}, {Ward-Thompson}, \&
  {Barsony}}]{Andre2000}
{Andr\'e}, P., {Ward-Thompson}, D., \& {Barsony}, M. 2000, Protostars and
  Planets IV, 59

\bibitem[{{Compi{\`e}gne} {et~al.}(2011){Compi{\`e}gne}, {Verstraete}, {Jones},
  {Bernard}, {Boulanger}, {Flagey}, {Le Bourlot}, {Paradis}, \&
  {Ysard}}]{Compiegne2011}
{Compi{\`e}gne}, M., {Verstraete}, L., {Jones}, A., {et~al.} 2011, \aap, 525,
  A103

\bibitem[{{Cook} \& {Stefanski}(1994)}]{CookStefanski1994}
{Cook}, J. \& {Stefanski}, L. 1994, Journal of the Americal Statistical
  Association, 89, 428

\bibitem[{{D{\'e}sert} {et~al.}(2008){D{\'e}sert}, {Mac{\'{\i}}as-P{\'e}rez},
  {Mayet}, {Giardino}, {Renault}, {Aumont}, {Beno{\^i}t}, {Bernard},
  {Ponthieu}, \& {Tristram}}]{Desert2008}
{D{\'e}sert}, F., {Mac{\'{\i}}as-P{\'e}rez}, J.~F., {Mayet}, F., {et~al.} 2008,
  A\&A, 481, 411

\bibitem[{{Draine}(2003)}]{Draine2003}
{Draine}, B.~T. 2003, \apj, 598, 1017

\bibitem[{{Dupac} {et~al.}(2003){Dupac}, {Bernard}, {Boudet}, {Giard},
  {Lamarre}, {M{\'e}ny}, {Pajot}, {Ristorcelli}, {Serra}, {Stepnik}, \&
  {Torre}}]{Dupac2003}
{Dupac}, X., {Bernard}, J., {Boudet}, N., {et~al.} 2003, A\&A, 404, L11

\bibitem[{{Enoch} {et~al.}(2007){Enoch}, {Glenn}, {Evans}, {Sargent}, {Young},
  \& {Huard}}]{Enoch2007}
{Enoch}, M.~L., {Glenn}, J., {Evans}, II, N.~J., {et~al.} 2007, \apj, 666, 982

\bibitem[{{Evans} {et~al.}(2001){Evans}, {Rawlings}, {Shirley}, \&
  {Mundy}}]{Evans2001}
{Evans}, II, N.~J., {Rawlings}, J.~M.~C., {Shirley}, Y.~L., \& {Mundy}, L.~G.
  2001, \apj, 557, 193

\bibitem[{{Goodman} {et~al.}(2009){Goodman}, {Pineda}, \&
  {Schnee}}]{Goodman2009}
{Goodman}, A.~A., {Pineda}, J.~E., \& {Schnee}, S.~L. 2009, \apj, 692, 91

\bibitem[{{Jones} \& {Nuth}(2011)}]{Jones2011}
{Jones}, A.~P. \& {Nuth}, J.~A. 2011, \aap, 530, A44

\bibitem[{{Juvela} {et~al.}(2012{\natexlab{a}}){Juvela}, {Malinen}, \&
  {Lunttila}}]{Juvela2012_mhdfil}
{Juvela}, M., {Malinen}, J., \& {Lunttila}, T. 2012{\natexlab{a}}, \aap, 544,
  A141

\bibitem[{{Juvela} {et~al.}(2008){Juvela}, {Pelkonen}, {Padoan}, \&
  {Mattila}}]{Juvela2008}
{Juvela}, M., {Pelkonen}, V.-M., {Padoan}, P., \& {Mattila}, K. 2008, \aap,
  480, 445

\bibitem[{{Juvela} {et~al.}(2009){Juvela}, {Pelkonen}, \&
  {Porceddu}}]{Juvela2009}
{Juvela}, M., {Pelkonen}, V.-M., \& {Porceddu}, S. 2009, \aap, 505, 663

\bibitem[{{Juvela} {et~al.}(2012{\natexlab{b}}){Juvela}, {Ristorcelli},
  {Pagani}, {Doi}, {Pelkonen}, {Marshall}, {Bernard}, {Falgarone}, {Malinen},
  {Marton}, {McGehee}, {Montier}, {Motte}, {Paladini}, {T{\'o}th}, {Ysard},
  {Zahorecz}, \& {Zavagno}}]{GCC_III}
{Juvela}, M., {Ristorcelli}, I., {Pagani}, L., {et~al.} 2012{\natexlab{b}},
  \aap, 541, A12

\bibitem[{{Juvela} {et~al.}(2011){Juvela}, {Ristorcelli}, {Pelkonen},
  {Marshall}, {Montier}, {Bernard}, {Paladini}, {Lunttila}, {Abergel},
  {Andr{\'e}}, {Dickinson}, {Dupac}, {Malinen}, {Martin}, {McGehee}, {Pagani},
  {Ysard}, \& {Zavagno}}]{GCC_II}
{Juvela}, M., {Ristorcelli}, I., {Pelkonen}, V.-M., {et~al.} 2011, \aap, 527,
  A111+

\bibitem[{{Juvela} \& {Ysard}(2012{\natexlab{a}})}]{Juvela2012_chi2}
{Juvela}, M. \& {Ysard}, N. 2012{\natexlab{a}}, \aap, 541, A33

\bibitem[{{Juvela} \& {Ysard}(2012{\natexlab{b}})}]{Juvela2012_TB}
{Juvela}, M. \& {Ysard}, N. 2012{\natexlab{b}}, \aap, 539, A71

\bibitem[{{Kelly} {et~al.}(2012){Kelly}, {Shetty}, {Stutz}, {Kauffmann},
  {Goodman}, \& {Launhardt}}]{Kelly2012}
{Kelly}, B.~C., {Shetty}, R., {Stutz}, A.~M., {et~al.} 2012, \apj, 752, 55

\bibitem[{{Lombardi} {et~al.}(2006){Lombardi}, {Alves}, \&
  {Lada}}]{Lombardi2006}
{Lombardi}, M., {Alves}, J., \& {Lada}, C.~J. 2006, \aap, 454, 781

\bibitem[{{Lunttila} \& {Juvela}(2012)}]{Lunttila2012}
{Lunttila}, T. \& {Juvela}, M. 2012, \aap, 544, A52

\bibitem[{{Malinen} {et~al.}(2011){Malinen}, {Juvela}, {Collins}, {Lunttila},
  \& {Padoan}}]{Malinen2011}
{Malinen}, J., {Juvela}, M., {Collins}, D.~C., {Lunttila}, T., \& {Padoan}, P.
  2011, \aap, 530, A101+

\bibitem[{{Meny} {et~al.}(2007){Meny}, {Gromov}, {Boudet}, {Bernard},
  {Paradis}, \& {Nayral}}]{Meny2007}
{Meny}, C., {Gromov}, V., {Boudet}, N., {et~al.} 2007, \aap, 468, 171

\bibitem[{{Motte} {et~al.}(1998){Motte}, {Andre}, \& {Neri}}]{Motte1998}
{Motte}, F., {Andre}, P., \& {Neri}, R. 1998, A\&A, 336, 150

\bibitem[{{Ormel} {et~al.}(2011){Ormel}, {Min}, {Tielens}, {Dominik}, \&
  {Paszun}}]{Ormel2011}
{Ormel}, C.~W., {Min}, M., {Tielens}, A.~G.~G.~M., {Dominik}, C., \& {Paszun},
  D. 2011, \aap, 532, A43

\bibitem[{{Ossenkopf} \& {Henning}(1994)}]{Ossenkopf1994}
{Ossenkopf}, V. \& {Henning}, T. 1994, A\&A, 291, 943

\bibitem[{{Padoan} \& {Nordlund}(2011)}]{PadoanNordlund2011}
{Padoan}, P. \& {Nordlund}, {\AA}. 2011, \apj, 730, 40

\bibitem[{{Paradis} {et~al.}(2010){Paradis}, {Veneziani}, {Noriega-Crespo},
  {Paladini}, {Piacentini}, {Bernard}, {de Bernardis}, {Calzoletti},
  {Faustini}, {Martin}, {Masi}, {Montier}, {Natoli}, {Ristorcelli}, {Thompson},
  {Traficante}, \& {Molinari}}]{Paradis2010}
{Paradis}, D., {Veneziani}, M., {Noriega-Crespo}, A., {et~al.} 2010, \aap, 520,
  L8

\bibitem[{{Pilbratt} {et~al.}(2010){Pilbratt}, {Riedinger}, {Passvogel},
  {Crone}, {Doyle}, {Gageur}, {Heras}, {Jewell}, {Metcalfe}, {Ott}, \&
  {Schmidt}}]{Pilbratt2010}
{Pilbratt}, G.~L., {Riedinger}, J.~R., {Passvogel}, T., {et~al.} 2010, \aap,
  518, L1

\bibitem[{{Planck Collaboration} {et~al.}(2011){Planck Collaboration}, {Ade},
  {Aghanim}, {Arnaud}, {Ashdown}, {Aumont}, {Baccigalupi}, {Balbi}, {Banday},
  {Barreiro}, \& et~al.}]{PlanckI}
{Planck Collaboration}, {Ade}, P.~A.~R., {Aghanim}, N., {et~al.} 2011, \aap,
  536, A23

\bibitem[{{Roussel}(2012)}]{Roussel2012}
{Roussel}, H. 2012, ArXiv e-prints

\bibitem[{{Shetty} {et~al.}(2009{\natexlab{a}}){Shetty}, {Kauffmann}, {Schnee},
  \& {Goodman}}]{Shetty2009b}
{Shetty}, R., {Kauffmann}, J., {Schnee}, S., \& {Goodman}, A.~A.
  2009{\natexlab{a}}, \apj, 696, 676

\bibitem[{{Shetty} {et~al.}(2009{\natexlab{b}}){Shetty}, {Kauffmann}, {Schnee},
  {Goodman}, \& {Ercolano}}]{Shetty2009a}
{Shetty}, R., {Kauffmann}, J., {Schnee}, S., {Goodman}, A.~A., \& {Ercolano},
  B. 2009{\natexlab{b}}, \apj, 696, 2234

\bibitem[{{Stamatellos} \& {Whitworth}(2003)}]{StamatellosWhitworth2003}
{Stamatellos}, D. \& {Whitworth}, A.~P. 2003, \aap, 407, 941

\bibitem[{{Stepnik} {et~al.}(2003){Stepnik}, {Abergel}, {Bernard}, {Boulanger},
  {Cambr{\'e}sy}, {Giard}, {Jones}, {Lagache}, {Lamarre}, {Meny}, {Pajot}, {Le
  Peintre}, {Ristorcelli}, {Serra}, \& {Torre}}]{Stepnik2003}
{Stepnik}, B., {Abergel}, A., {Bernard}, J., {et~al.} 2003, \aap, 398, 551

\bibitem[{{Tauber} {et~al.}(2010){Tauber}, {Mandolesi}, {Puget}, {Banos},
  {Bersanelli}, {Bouchet}, {Butler}, {Charra}, {Crone}, {Dodsworth}, \&
  et~al.}]{Tauber2010}
{Tauber}, J.~A., {Mandolesi}, N., {Puget}, J., {et~al.} 2010, \aap, 520, A1

\bibitem[{{Veneziani} {et~al.}(2010){Veneziani}, {Ade}, {Bock}, {Boscaleri},
  {Crill}, {de Bernardis}, {De Gasperis}, {de Oliveira-Costa}, {De Troia}, {Di
  Stefano}, {Ganga}, {Jones}, {Kisner}, {Lange}, {MacTavish}, {Masi},
  {Mauskopf}, {Montroy}, {Natoli}, {Netterfield}, {Pascale}, {Piacentini},
  {Pietrobon}, {Polenta}, {Ricciardi}, {Romeo}, \& {Ruhl}}]{Veneziani2010}
{Veneziani}, M., {Ade}, P.~A.~R., {Bock}, J.~J., {et~al.} 2010, \apj, 713, 959

\bibitem[{{Ysard} {et~al.}(2012){Ysard}, {Juvela}, {Demyk}, {Guillet},
  {Abergel}, {Bernard}, {Malinen}, {M{\'e}ny}, {Montier}, {Paradis},
  {Ristorcelli}, \& {Verstraete}}]{YsardJuvela2012}
{Ysard}, N., {Juvela}, M., {Demyk}, K., {et~al.} 2012, \aap, 542, A21

\end{thebibliography}

\end{document}